\documentclass[prd,showpacs,groupedaddress,superscriptaddress,twocolumn,
10pt,nofootinbib,floatfix,preprintnumbers]{revtex4-1}
\usepackage{hyperref,amssymb,amsmath,graphicx,xcolor,cancel}

\DeclareMathOperator{\Tr}{Tr}

\newcommand{\slashed}{/\!\!\!}
\makeatletter
\let\@bibitemShut=\@empty
\makeatother

\begin{document}

\title{Nucleon Charges and Electromagnetic Form Factors from 2+1+1-Flavor Lattice QCD}

\author{Tanmoy Bhattacharya}
\affiliation{Los Alamos National Laboratory, Theoretical Division T-2, Los Alamos, NM 87545}

\author{Saul D. Cohen}
\affiliation{Department of Physics, University of Washington, Seattle, WA 98195-1560}

\author{Rajan Gupta}
\affiliation{Los Alamos National Laboratory, Theoretical Division T-2, Los Alamos, NM 87545}

\author{Anosh Joseph}
\affiliation{Los Alamos National Laboratory, Theoretical Division T-2, Los Alamos, NM 87545}

\author{Huey-Wen Lin}
\affiliation{Department of Physics, University of Washington, Seattle, WA 98195-1560}

\author{Boram Yoon}
\affiliation{Los Alamos National Laboratory, Theoretical Division T-2, Los Alamos, NM 87545}

\collaboration{Precision Neutron-Decay Matrix Elements (PNDME) Collaboration}

\date{\today}

\preprint{LA-UR-13-24606}
\preprint{NT@UW-13-23}


\begin{abstract}

We present lattice-QCD results on the nucleon isovector axial, scalar
and tensor charges, the isovector electromagnetic Dirac and Pauli form
factors, and the connected parts of the isoscalar charges. The
calculations have been done using two ensembles of HISQ lattices
generated by the MILC Collaboration with 2+1+1 dynamical flavors at a
lattice spacing of $0.12$~fm and with light-quark masses corresponding
to pions with masses $310$ and $220$~MeV. We perform a systematic
study including excited-state degrees of freedom and examine the
dependence of the extracted nucleon matrix elements on source-sink
separation.  This study demonstrates with high-statistics data that
including excited-state contributions and generating data at multiple
separations is necessary to remove contamination that would otherwise
lead to systematic error.  We also determine the renormalization
constants of the associated quark bilinear operators in the RI-sMOM
scheme and make comparisons of our renormalized results with previous
dynamical-lattice calculations.
\end{abstract}

\maketitle

\section{Introduction}
\label{sec:intro}

Precision measurements of the properties of neutrons provide an
opportunity to determine a fundamental parameter of the Standard
Model (SM), to compute many parameters of effective theories in nuclear physics
and to explore novel physics at the TeV scale. Decays
of neutrons provide one of the best determinations of the CKM quark-mixing
parameter $V_{ud}$ and the nucleon axial charge $g_A$. Calculations of nucleon matrix
elements of bilinear quark operators yield a variety of interesting physical quantities, including
the nucleon charges such as $g_{A,S,T}$, the nucleon $\sigma$-term and strangeness, and the
electromagnetic form factors.
In addition to these quantities, isoscalar bilinear matrix elements can
be related to measurements of the neutron electric dipole moment (nEDM), which
will shed light on CP violation in and beyond the Standard Model (BSM).
In this paper we report a lattice-QCD (LQCD)
calculation of the isovector charges $g_A$, $g_S$ and $g_T$,
the connected-diagram part of the isoscalar charges
and the isovector electric and magnetic radii, extracted from the
electromagnetic form factors. The axial charge $g_A$ is a key
parameter in nuclear physics, while estimates of $g_S$ and $g_T$
are needed to constrain possible scalar and tensor interactions at the
TeV scale~\cite{Bhattacharya:2011qm}.

The nucleon isovector axial charge $g_A$ is a key parameter in the description
of nucleon structure, since it encapsulates the interaction of the charged
axial current with the nucleon. For example, it affects the rate of
proton-proton fusion, which is the first step in the thermonuclear
reaction chains that power low-mass hydrogen-burning stars like the
Sun; and $g_A$ is central to the extraction and phenomenology of the CKM
matrix element $V_{ud}$. Presently, $g_A$ is best known from the
experimental measurement of neutron beta decay using polarized
ultracold neutrons by the UCNA collaboration~\cite{Plaster:2008si}, which
dominates the PDG average with uncertainty at the 0.2\%
level~\cite{Beringer:1900zz}. (See the figure in Sec.~\ref{sec:charges}
for the collected experimental $g_A$ measurements
used in PDG2012, the recently updated UCNA
number $1.2756(30)$~\cite{Mendenhall:2012tz}
and a recent result from Perkeo~II $1.2761^{+14}_{-17}$~\cite{Mund:2012fq}.)

The nucleon scalar and tensor charges have not, until very recently,
been well studied since the contributions of effective scalar and
tensor interactions in the SM are small, at the $10^{-3}$ level,
and still below the current experimental limits. In many extensions
to the SM (e.g. supersymmetry), novel scalar and/or tensor
interactions arise via exchanges in either the $s$ or $t$ channels or
through loop effects, and these can also contribute at the $10^{-3}$
level to neutron decay. Since the SM contributions are known to high
precision, $10^{-5}$, one has the opportunity to measure these scalar
and tensor couplings in neutron beta decay experiments with sufficient
precision to isolate the BSM from the SM contributions. The current
status of the theory and experimental measurements is summarized in
Ref.~\cite{Bhattacharya:2011qm}, in which it was shown that, assuming
that planned experiments achieve $10^{-3}$ sensitivity, to constrain
new physics at the TeV scale estimates of $g_S$ and $g_T$ with
10--15\% accuracy are needed. A number of such experiments are being
developed at Los Alamos (UCNB~\cite{WilburnUCNB} and
UCNb~\cite{UCNb}) and Oak Ridge National Lab
(Nab~\cite{Pocanic:2008pu,Nab:expt,Baessler:2012nc}), and they aim to make measurements
in the coming years.

Lattice QCD is the best method to obtain $g_S$ and $g_T$
with the desired precision. The calculation steps, conceptually and
procedurally, are the same as for the vector and axial charges. The
main hurdles are the statistical error in $g_S$, which are a factor of
about five larger than those in $g_A$ and $g_T$.

Lattice calculations of the properties of nucleons
are more difficult than those for mesons for a number of reasons.
First, the lowest excited state, the Roper $N(1440)$, lies relatively
close to the nucleon mass. Therefore, a creation operator that has substantial
overlap with excited states will require large values of Euclidean time
$t$ to reduce contributions of excited states to correlation functions.
Alternatively, the nucleon operators must be carefully crafted to maximize
overlap with the ground state.
Second, there is a signal-to-noise problem. In
Euclidean space, the signal-to-noise ratio in nucleon correlation
functions scales like $\exp((-M_N+3M_\pi/2)t)$. Since $M_N >
1.5M_\pi$, high statistics are needed to obtain a signal at large
$t$, where excited-state contributions have become negligible,
in order to extract the properties of the ground state. With finite
computer resources we must, therefore, trade off between statistical
and systematic error due to contamination from excited states. As a
result, much larger computational resources are needed for precision
studies of nucleons on the lattice than would be required for mesons.
Third, heavy-baryon chiral perturbation theory (HBXPT) is more difficult because of the nearby $\Delta$
resonance~\cite{Beane:2004rf}. As a result, there are several different HBXPT
expansions, and it is not a~priori clear how to
tell which will converge well for a given
observable. Consequently, although chiral perturbation theory has been a very
important tool for understanding the dependence of
meson observables on the light-quark masses in a lattice calculation,
it is not as useful for baryons. Thus, extrapolations of lattice
calculations from unphysically large light-quark ($u$ and $d$) masses to the physical point
have significant uncertainty. This problem is being
addressed gradually as simulations are being performed closer to
physical values of the light-quark masses.
Lastly, finite-volume effects are observed to be larger in baryon
correlation functions as compared to mesons. Historically, studies of finite-volume effects have
been carried out using mesonic observables, for which it has been
established empirically that finite-volume effects are negligible for
$M_\pi L > 4$. Since the generation of gauge ensembles is expensive
and driven by physics goals in the meson sector, only a
single ensemble of lattices with $M_\pi L > 4$ is usually generated
at a given lattice spacing. As a result, finite-volume effects are
less well understood in the baryon sector. A few studies suggest even larger values of $M_\pi L$ are needed to have this systematic under control~\cite{Hall:2013oga,Lin:2012ev,Renner:2010ks}.

In this work we present a detailed analysis of excited-state contamination in matrix elements
and results on the three isovector charges
$g_A$, $g_S$ and $g_T$, corresponding to the matrix elements of
isovector quark bilinear operators within the nucleon, and the Dirac
and Pauli (or the related electric and magnetic) charge radii
extracted from the corresponding electromagnetic form factors. We
also present the connected part of the isoscalar charges. All
calculations have been done at one lattice spacing, $a \approx 0.12$~fm,
so no extrapolation to the continuum limit is possible.
Ensembles of gauge configurations generated at two values of the light
quark mass, corresponding to $M_\pi \approx 310$ and $220$~MeV, have
been analyzed, so we will make some comments on quark-mass
dependence. Our main focus will be on understanding three issues:
statistical errors, excited-state contributions, and nonperturbative
calculations of the renormalization constants in the RI-sMOM scheme.

The paper is organized as follows.  In Sec.~\ref{sec:setup} we
describe the gauge ensembles and lattice parameters used in this
study.  Details of the calculations of the two- and three-point functions
are given in Sec.~\ref{sec:results}.  Analysis of the statistical
signal is presented in Sec.~\ref{sec:stat} and of the excited-state
contamination in Sec.~\ref{sec:excited}.
We discuss the calculation of the renormalization
constants in the RI-sMOM scheme in Sec.~\ref{sec:Zfac}.  The results for the charges and
comparison with other published
estimates are given in Secs.~\ref{sec:charges} and \ref{sec:isoscalar}.  The electromagnetic
form factors and charge radii are discussed in
Sec.~\ref{sec:formfactors}. Our conclusions are given in
Sec.~\ref{sec:end}.

\section{Lattice Parameters and Setup}
\label{sec:setup}

We analyze two ensembles of gauge configurations 
generated by the MILC collaboration~\cite{Bazavov:2012xda}
with $N_f=2+1+1$ flavors of highly improved staggered quarks
(HISQ)~\cite{Follana:2006rc,Follana:2004mg,Bazavov:2010ru,Bazavov:2009wm,Bazavov:2009jc,Bazavov:2009bb}
as described in Table~\ref{tab:param1}. The HISQ action, proposed by
HPQCD/UKQCD Collaboration~\cite{Follana:2006rc, Follana:2004mg},
has, among existing variations of staggered fermions,
at nonzero $a$ the smallest splittings between the
staggered ``tastes'' that become four degenerate flavors in the
continuum limit~\cite{Bazavov:2009wm, Bazavov:2011nk};
this leads to a significant reduction in the discretization errors
associated with the staggered action. The work
presented here is the first step in our analysis of HISQ sea-quark
ensembles of about 1000 configurations generated at three
lattice spacings $a\in \{0.12, 0.09, 0.06\}$~fm, two
light-quark masses corresponding to $M_\pi \approx 310$ and $220$~MeV and the strange and charm-quark masses set to their physical values. (The actual values of the lattice spacings we use in this study,
based on the data presented in Ref.~\cite{Bazavov:2012xda}, are
$a=0.120(1)$, $0.088(1)$ and $0.058(1)$~fm.)
Our goal is to perform a simultaneous continuum and chiral
extrapolation of physical quantities using these six ensembles. In
this paper, we focus on demonstrating control over systematic errors
associated with our lattice approximations using the two ensembles at
$a\approx 0.12$~fm.

Staggered-type fermions are notorious for their complications in
calculations involving baryons, especially of matrix elements.
Therefore, we use clover ($O(a)$-improved Wilson) fermion action in
the valence sector for our calculation of nucleon matrix
elements. Strictly speaking, such a mixed-action approach with HISQ
fermions for sea quarks and clover for valence quarks, results in a
non-unitary formulation. One consequence of this mixed-action approach
is the possibility of exceptional configurations. These are
configurations in which the spectrum of the clover Dirac matrix has
near-zero modes. Such configurations would have been suppressed if the
lattices had been generated with the same clover action, so their
presence is an artifact of using the mixed-action approach. Signatures
of such configurations, which manifest at sufficiently small quark
masses, include: (i) correlation functions calculated on them have
anomalously large values, thus biasing the ensemble average, and/or
(ii) the calculation of the inverse of the clover Dirac matrix fails
to converge due to poor condition number. Our approach to this problem
is empirical. Based on the two signatures listed above we have
determined that exceptional configurations are absent in the data on
$M_\pi\in \{220, 310\}$~MeV MILC ensembles~\cite{Bazavov:2012xda} at
0.12~fm, but the same is not the case for the ensemble at $M_\pi
\approx 135$~MeV. On these we find signatures of exceptional
configurations, and therefore do not analyze them. We expect that
exceptional configurations will be suppressed at finer lattice spacing,
and our ongoing analysis of an ensemble with $a=0.09$~fm and
$M_\pi=130$ MeV shows no exceptional configurations on the full set
(883 configurations). Thus, the mixed-action approach can be used at the
physical quark mass for lattice spacings of $0.09$~fm and smaller.

The mixed-action approach is used for two reasons. First, to calculate matrix
elements within baryon states, we need high-statistics analyses on
lattices with large volume ($M_\pi L > 4$). Second, to take the
continuum limit and to elucidate the dependence on pion mass
requires large ensembles at multiple pion masses
and lattice spacings. Because the generation of ensembles requires
very large computing resources sustained over many years, few
collaborations can meet these requirements. The MILC 2+1+1-flavor ensembles
are the only ones that satisfy these requirements that are
available to us. Since we are able to avoid exceptional
configurations, the mixed-action allows us to test new-physics ideas and
computational methods. In our mixed action
approach, lattice discretization errors are dominated by our Wilson-clover
action, which has not been fully $O(a)$ improved;
the the clover term has not been nonperturbatively tuned, but merely
set to the tadpole-improved
perturbative coefficient. Also, the operators used to calculate the matrix elements
and renormalization constants have not been improved. Thus, our
discretization errors start at $O(a)$, and we
must include a linear term in continuum extrapolation.

We use hypercubic (HYP) smearing~\cite{Hasenfratz:2001hp} of the gauge
links before inverting the clover Dirac matrix needed to construct
correlation functions~\cite{Bratt:2010jn,Lin:2007ap}. Using gauge
fields averaged over a hypercube reduces short-distance noise (lattice
artifacts) without changing long-distance physics. We observe this
improvement in the calculations of two and three-point correlation
functions. HYP smearing also modifies the discretization
artifacts appearing at high momentum in the calculation of the renormalization constants~\cite{Arthur:2013bqa}. We describe our strategy for
estimating the associated systematic uncertainty
in Sec.~\ref{sec:Zfac}. We find that even using
conservative error estimates, since
HYP smearing drives the renormalization constants close to the
tree-level value (unity), the uncertainty due to
renormalization constants in our preliminary
study~\cite{Bhattacharya:2011qm} is reduced.

Further details regarding the tuning of the valence clover action to
match the HISQ sea-quark action and issues regarding the mixed action
are discussed in Refs.~\cite{Briceno:2012wt, Briceno:2011cb,Lin:2011zzo, Gupta:2012rf, Bhattacharya:2012wk, Gupta:2012az}. In
Table~\ref{tab:param1}, we show the level of agreement between the pion
and nucleon masses calculated with the two actions.
Similar parameter choices for the same valence and sea-quark actions in the
light-quark sector are also used in a study of charmed-hadron physics
in Refs.~\cite{Briceno:2012wt, Briceno:2011cb}.

\begin{table*}
\begin{tabular}{|cccc|cccc|ccc|}
\hline
$\beta$ & $L^3\times T$ & $N_\text{cfgs}$ & $N_\text{props}$ & $(aM_\pi)_\text{sea}$ & $(aM_\pi)_\text{val}$  & $(M_\pi L)_\text{val}$ & $(M_\pi T)_\text{val}$ &  $(aM_N)_\text{sea}$  & $(aM_N)_\text{val}$ & $(aM_R)_\text{val}$ \\
\hline
$6.00$  &$24^3\times 64$ &1013&4052
&0.1893(1) &0.18947(30) &$4.6$&12.1  &  0.708(8)   & 0.6689(65)  & 1.46(15)   \\\hline
$6.00$ &$32^3\times 64$ &958&3832
&0.13407(6) & 0.13718(33) & $4.4$ & 8.8 & 0.647(6)   & 0.6255(72)  & 1.45(9)    \\\hline
\end{tabular}
\caption{Details of the two ensembles analyzed and lattice parameters
  used in this study. The subscript ``sea'' labels the masses of the
  Goldstone pseudoscalar meson and nucleon calculated using the HISQ
  on HISQ action~\cite{Bazavov:2012xda}, while the subscript ``val''
  labels the masses calculated with the valence clover fermions on the
  HISQ lattices. The ``sea'' masses have a single statistical
  uncertainty, while the valence masses include statistical and
  systematic uncertainty due to fitting-window selection added in
  quadrature. We also list the
  spatial ($L$) and temporal lattice extents ($T$) in lattice units,
  the value of $ M_\pi L$, $M_\pi T$, the number of configurations
  analyzed, and the total number of measurements ($N_\text{props}$)
  performed on each ensemble. Note that although we call our fitted excited-state nucleon mass 
  $M_R$ the Roper mass, it requires study of higher excited states and the volume dependence of the correlator
  to properly distinguish a true Roper resonance from a scattering state.}
\label{tab:param1}
\end{table*}

\section{Lattice Methodology}
\label{sec:results}

In this section we describe the lattice calculation of two- and
three-point correlation functions. After establishing the notation and
methodology in Secs.~\ref{sec:two-point} and \ref{sec:three-point}, we
discuss the statistical errors in Sec.~\ref{sec:stat} and our
understanding and mitigation of excited-state contamination in the
extraction of the ground-state matrix elements in
Sec.~\ref{sec:excited}.

\subsection{Two-Point Correlators}
\label{sec:two-point}

The correlation functions with the quantum numbers of the spin-1/2 nucleon are
constructed using the baryonic interpolating operator
\begin{equation}
 \chi^N (x) = \epsilon^{abc} [q_1^{a\top}(x)C\gamma_5q_2^b(x)]q_1^c(x),
 \label{eq:lat_B-op}
\end{equation}
where $C$ is the charge-conjugation matrix $i \gamma_4 \gamma_2$, $\{a,b,c\}$ are color indices, $\epsilon$ is the antisymmetric tensor and $q_1$ and $q_2$ are
one of the two quarks $\{u,d\}$. For example, in the case of the proton,
we want $q_1=u$ and $q_2=d$. Two-point correlators are derived from
these interpolating fields as
\begin{equation}
{\cal C}^{(2)}_{AB}(t_f,t_i;\vec{p}) =
   \sum_{\vec{x}} e^{i\vec{p}\cdot\vec{x}}
   \langle {\cal P} \chi^N_A(\vec{x},t_f)\chi^N_B(\vec{0},t_i)^\dagger\rangle ,
\label{eq:two-pt-correlators1}
\end{equation}
where $\vec{p}$ is the baryon momentum and
${\cal P} =\frac{1}{4}(1+\gamma_4)(1+i\gamma_5\gamma_3)$ is the spin projection. $A$ and $B$
label the smearing parameters used for the source and sink operators
as discussed below. Eq.~\ref{eq:two-pt-correlators1} can be decomposed
in terms of energy eigenstates:
\begin{equation}
{\cal C}^{(2)}_{AB}(t_f,t_i;\vec{p}) =
   \sum_n \frac{E_n(\vec{p})+M_n}{2E_n(\vec{p})} {\cal A}_{n,A} {\cal A}_{n,B}
          e^{-E_n(\vec{p}) (t_f-t_i)},
 \label{eq:two-pt-correlators2}
\end{equation}
where $n$ runs over all energy eigenstates that couple to the
operator defined in Eq.~\ref{eq:lat_B-op} with amplitude ${\cal A}_{n,A}$
for smearing parameter $A$.
The normalization of these states is defined as $\langle 0
|(\chi^N)^\dagger|p,s\rangle = {\cal A} u_N(\vec p,s)$ with the spinors in Euclidean space satisfying
\begin{equation}
\sum_{s} u_N(\vec p,s) \bar{u}_N(\vec{p},s) =
   \frac{E(\vec{p})\gamma_4-i\vec\gamma\cdot \vec{p} + M}{2 E(\vec{p})}.
\label{eq:spinor}
\end{equation}

\begin{figure}
\includegraphics[width=0.45\textwidth]{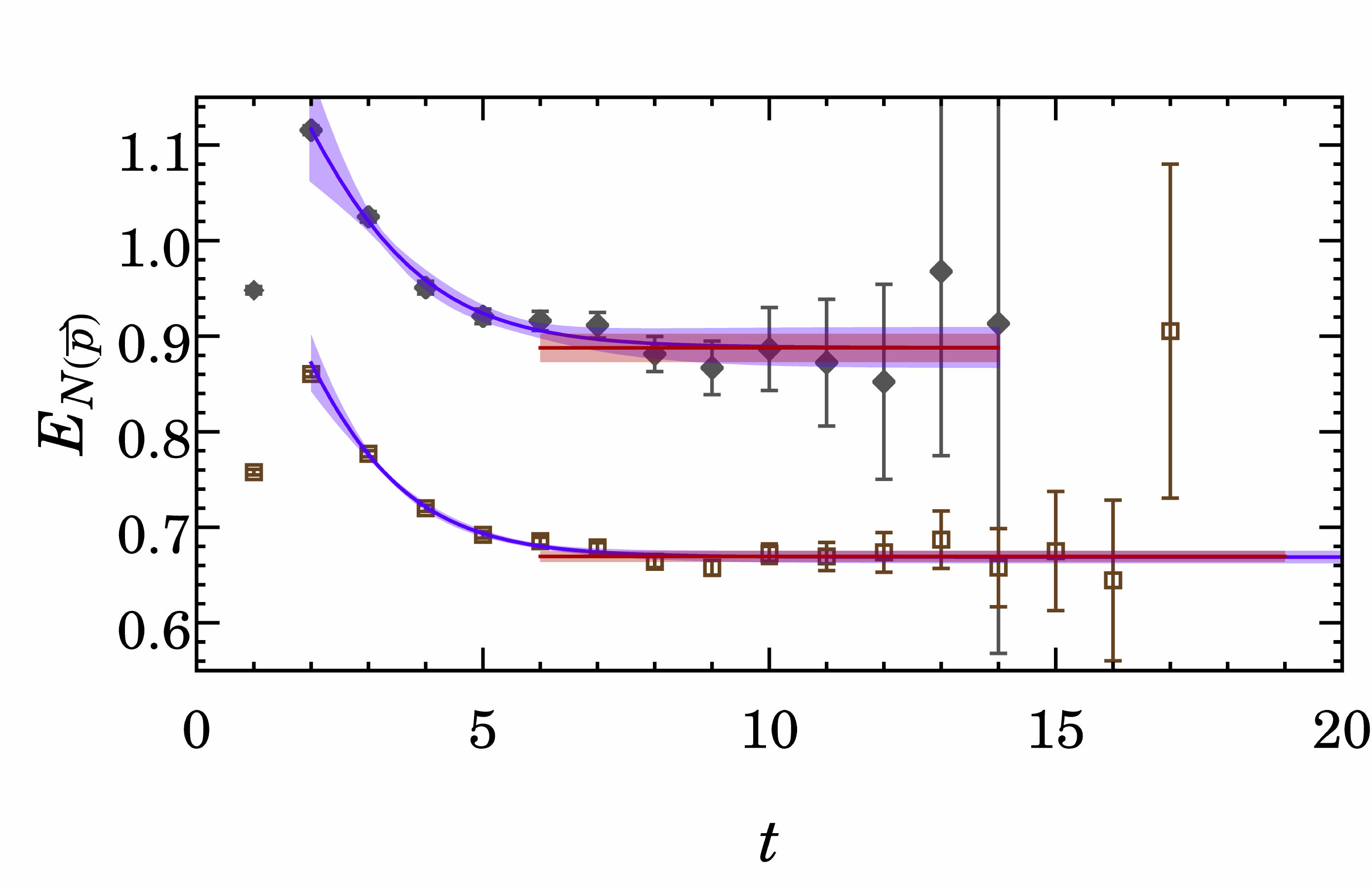}
\includegraphics[width=0.45\textwidth]{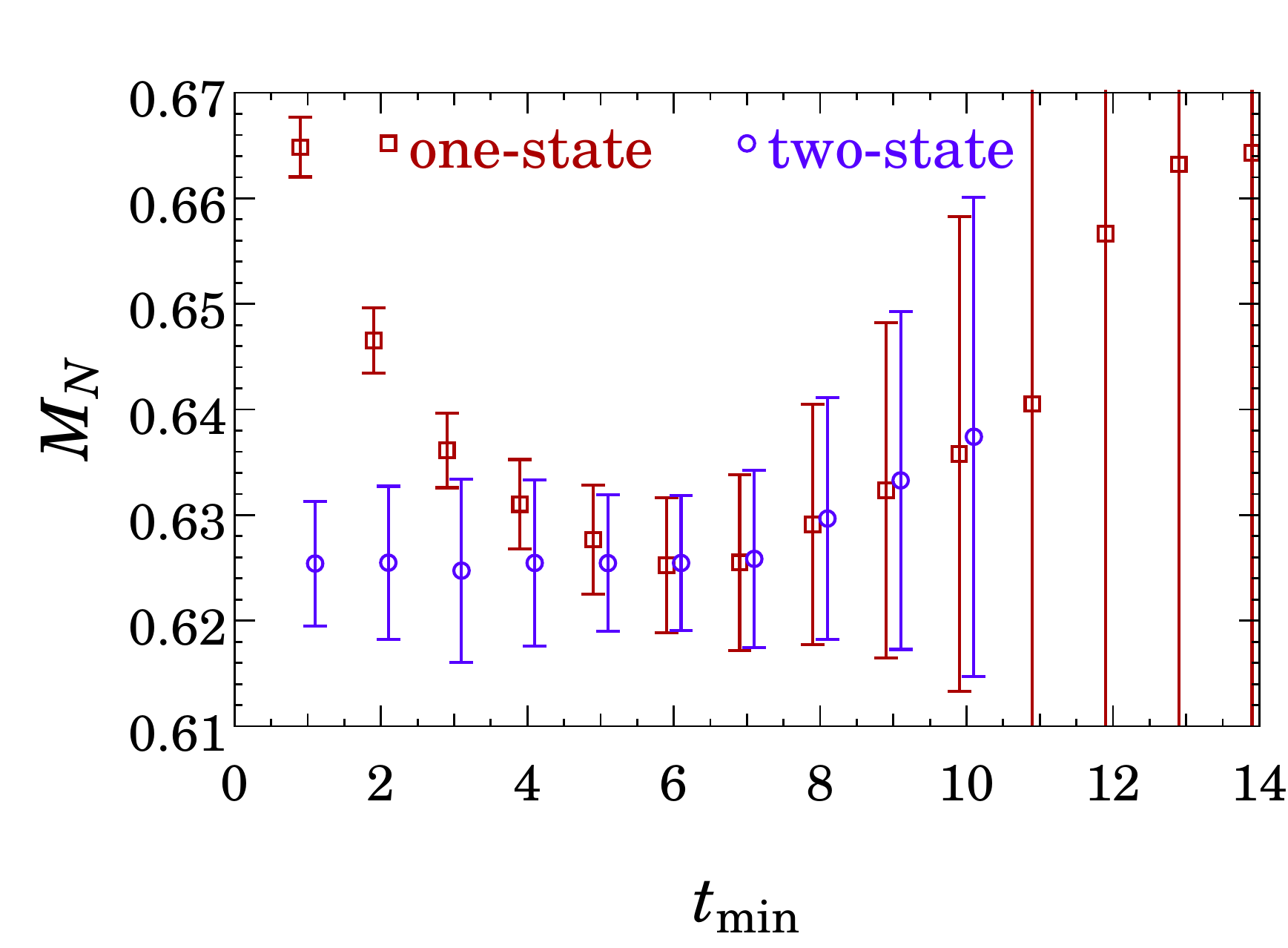}
\caption{(top) Nucleon effective-mass plot with one- and
  two-state fits for the 310-MeV ensemble described in
  Table~\ref{tab:param1}. The data are for momenta $\vec{p}=0$ (squares) and
  $|\vec{p}|^2=5 \left(\frac{2\pi}{La}\right)^2$ (diamonds). Quark propagators
  with the clover action were calculated using Gaussian-smearing sources with
  parameters $\{\sigma,n_\text{KG}\}=\{5.5,70\}$ as described in the text.
  (bottom) A comparison of fitted values of nucleon mass with one- and
  two-state fits as functions of $t_\text{min}$, the starting value of $t$
  used in the fits. The data are for the $M_\pi \approx 220$~MeV
  ensemble described in Table~\ref{tab:param1}. The two
  fits agree for $t_\text{min} \ge 6$, and the two-state fit yields a consistent ground-state mass for all $t_\text{min}>0$.
   }
\label{fig:mNeff}
\end{figure}

To construct correlation functions, we generate valence clover quark
propagators with a gauge-invariant Gaussian-smeared source centered at
$x$, the point at which the nucleon operator is defined in
Eq.~\ref{eq:lat_B-op}. Smearing is done using a fixed number, $n_\text{KG}$,
of applications of the Klein-Gordon operator with coefficient
$\sigma$.  The ideal smearing for calculating the ground-state
zero-momentum nucleon mass and matrix elements leading to $g_{A,S,T}$
is one that minimizes overlap with excited states. An effective-mass
plot with the results of one- and two-state fits for the zero-momentum
nucleon is given on the top of Fig.~\ref{fig:mNeff} with diagonal
Gaussian smearing $A=B$.  For Gaussian smearing parameters of
$\{\sigma,n_\text{KG}\}=\{5.5,70\}$ (using the conventions found in
Chroma~\cite{Edwards:2004sx}), we find that the excited-state signal
dies out around $t=6$ ($\approx 0.7$~fm), giving enough data points to
extract ground-state and first-excited masses and amplitudes as
shown. The bottom of Fig.~\ref{fig:mNeff} shows that the extracted
ground-state mass agrees between the two-state and one-state fits when
one-state fits are constrained to $t \ge 6$, and the two-state fit
gives a consistent ground-state mass for all fitting windows.  Our
estimate of the masses of the ground and first-excited states for each
ensemble are given in Table~\ref{tab:param1}.

One challenge in the selection of smearing parameters is the need to
simultaneously improve the signal in states at nonzero momentum, as high as say
$|\vec{p}|^2=5$ in units of $\frac{2\pi}{La}$, since these are needed
to study form factors. We find the signal in our smeared correlator
deteriorates significantly compared to smearings with smaller
radii. This is not surprising, since high-momentum states are expected
to have a smaller overlap with a broadly smeared source.  Thus, to
improve the quality of the signal at higher momenta, we need to
produce multiple correlators with different smearings and in each case
explicitly subtract any excited-state contributions to obtain results
for the ground state. The choice of smearing in this study was driven
by improving results for $g_{A,S,T}$. We, therefore, used diagonal
Gaussian-smeared sources $A=B$ with a single smearing $\{5.5,70\}$,
optimized to improve the signal in the zero-momentum two-point nucleon
correlator.

To simplify notation, we drop the subscripts defining the smearing,
$A$ and $B$, in all further discussions. The masses of the ground and
first-excited state will be labeled as $M_0$ and $M_1$ and
the corresponding amplitudes with which they couple to the operator
defined in Eq.~\ref{eq:lat_B-op} as ${\cal A}_0$ and
${\cal A}_1$. These masses and amplitudes are needed as inputs to
extract the charges and form factors from three-point correlators. Our
final results for these quantities are obtained by applying a
fit to the smeared-smeared zero-momentum
correlators, keeping only two states in Eq.~\ref{eq:two-pt-correlators2}:
\begin{equation}
{\cal C}^{(2)}(t_f,t_i;\vec{p})= {|{\cal A}_0|}^2 e^{-M_0 (t_f-t_i)} + {|{\cal A}_1|}^2 e^{-M_1 (t_f-t_i)} \ .
\label{eq:two-pt-correlators3}
\end{equation}

\subsection{Three-Point Correlators}
\label{sec:three-point}

To calculate the nucleon matrix elements (such as isovector charges or
electromagnetic form factors), we first calculate the matrix element
of general form $\langle \chi^N (\vec{p}_f) | O_\Gamma | \chi^N
(\vec{p}_i) \rangle$, where $O_\Gamma$ is $V^\mu=\overline{u}\gamma_\mu
d$ for the isovector vector current,
$A^\mu=\overline{u}\gamma_\mu \gamma_5 d$ for isovector axial current, etc., and
$\vec{p}_{\{i,f\}}$ are the initial and final nucleon momenta.
Such a matrix element is extracted from an appropriate three-point
correlation function after Fourier transforming out the spatial
dependence and projecting on the baryonic spin, leaving a time-dependent
three-point correlator of the form
\begin{multline}\label{eq:general-three-point}
{\cal C}^{(3),T}_{\Gamma}(t_i,t,t_f;\vec{p}_i,\vec{p}_f) = \\
Z_\Gamma \sum_{n,n^\prime} f_{n,n^\prime}
\sum_{s,s^\prime}
T_{\alpha\beta}\, u_{n^\prime}^\beta(\vec{p}_f,s^\prime) \times \\
\langle N_{n^\prime}(\vec{p}_f,s^\prime)\left|O_\Gamma\right|N_n(\vec{p}_i,s)\rangle\overline{u}_n^\alpha(\vec{p}_i,s),
\end{multline}
where $f_{n,n^\prime}$
contains kinematic factors involving the energies $E_n$ and amplitudes
${\cal A}_n$ between the creation and annihilation operators and the corresponding states.
The latter are obtained from the analysis of the two-point correlators
with $n$ and $n^\prime$ labeling the different energy
states. $Z_\Gamma$ is the operator renormalization constant which is
determined nonperturbatively in this work; see Sec.~\ref{sec:Zfac}. The projection $T$ used
is $T_\text{mix}=\frac{1}{4}(1+\gamma_4)(1+i\gamma_5\gamma_3)$.

In this work we are interested in only the ground-state matrix element
with $n=n^\prime=0$. The parameter of interest in quantifying excited-state contamination, discussed in Sec.~\ref{sec:excited}, is the
source-sink separation ($t_f-t_i$). In this study it is varied between
8 and 12 time slices in lattice units, which in physical units
corresponds to source-sink separations between about 0.96 and 1.44~fm.
By fitting the time dependence of the three-point correlators to the
form of Eq.~\ref{eq:general-three-point} with $n$ and $n^\prime$ restricted to
0 and 1, we isolate the matrix elements in the ground state from those
in the first excited state as shown in Sec.~\ref{sec:excited}.

To study electromagnetic form factors, we use one final momentum
($\vec{p}_f=\frac{2\pi}{L a}\{0,0,0\}$) and vary the initial
momenta over all $\vec{p}_i=\frac{2\pi}{L a}\{n_x,n_y,n_z\}$ with
integer $n_{x,y,z}$ such that $n_x^2+n_y^2+n_z^2 \leq 5$. For all other charges
we project onto $\vec{p}_i=0$ by inserting the operator at zero momentum.

\subsection{Statistics}
\label{sec:stat}

The MILC Collaboration has produced ensembles of roughly 5500
trajectories of 2+1+1-flavor HISQ lattices at the two quark
masses. The ensembles at $a\approx 0.12$~fm are described
in Table~\ref{tab:param1}. Five hundred trajectories are discarded
for thermalization, which is somewhat more conservative than the 300 discarded by MILC. Correlators are then calculated on configurations separated by
five trajectories. On each configuration, we use four smeared sources,
displaced both in time and space directions to reduce correlations. Furthermore, two sets of these four source points, again
maximally separated in space and time directions, are used on each
alternate configuration to reduce correlations. To evaluate the
statistical significance of the data, in addition to the full set, we
also analyze the roughly 500 configurations with each of these two
sets of sources. We verify that the two sets give compatible results
and the errors are roughly $\sqrt 2$ larger compared to the full
set.

\section{Excited-State Contamination}
\label{sec:excited}

All observables reported in this paper (charges, charge radii,
form factors) need to be calculated between ground-state nucleons. The
operators used to create and annihilate the states, defined in
Eq.~\ref{eq:lat_B-op}, however, couple to the nucleon and all its
radially excited states.
There are two possible ways to reduce contributions
from excited states: by reducing the overlap of the interpolating
operator with the excited states and by increasing the time separation
$t_\text{sep}=t_f-t_i$ between the source and sink to exponentially suppress
excited-state contamination. As discussed above, we use sources and
sinks with one fixed smearing size that improves overlap with the
ground state; nevertheless, the two-point correlator shows significant
excited-state contribution extending to $t=5$. Thus, using single-state analysis on the three-point correlator is likely to be problematic for small $t_\text{sep}$. Statistics limit the
upper value of $t_\text{sep}$ that can be explored, and we find that the
signal degrades very significantly by $t_\text{sep} = 12$. We, therefore,
investigate up to five time separations between $t_\text{sep}=8$ and 12 to quantify the
excited-state contamination as discussed below.

We consider the leading excited-state contamination mass $M_1$ and its coupling 
to our operator with amplitude ${\cal A}_1$. We can write
the three-point function with source shifted to $t_i=0$, operator insertion at
$t=t$ and sink at $t_f= t_\text{sep}$ as
\begin{align}
{\cal C}^{(3),T}_{\Gamma}&(t_i,t,t_f;\vec{p}_i,\vec{p}_f) \approx \nonumber\\
      & |{\cal A}_0|^2 \langle 0 | O_\Gamma | 0 \rangle  e^{-M_0 (t_f-t_i)} +{}\nonumber\\
      & |{\cal A}_1|^2 \langle 1 | O_\Gamma | 1 \rangle  e^{-M_1 (t_f-t_i)} +{}\nonumber\\
      & {\cal A}_0{\cal A}_1^* \langle 0 | O_\Gamma | 1 \rangle  e^{-M_0 (t-t_i)} e^{-M_1 (t_f-t)} +{}\nonumber\\
      & {\cal A}_0^*{\cal A}_1 \langle 1 | O_\Gamma | 0 \rangle  e^{-M_1 (t-t_i)} e^{-M_0 (t_f-t)} ,
\label{eq:three-pt}
\end{align}
where $\langle n^\prime | O_\Gamma | n \rangle$ is an abbreviation for $\langle N_{n^\prime}(\vec{p}_f,s^\prime)\left|O_\Gamma\right|N_n(\vec{p}_i,s)\rangle$. To extract $\langle 0 | O_\Gamma | 0
\rangle$ from the two- and three-point functions we make the following
different kinds of fits. In each case, we apply a nonlinear
least-square fitter that automatically selects a fit range appropriate to
the form used. For each form on each correlator, the fit range is expanded
as long as the quality of the fit (in terms of uncorrelated $\chi^2/\text{dof}$) does not
sharply decline.
\begin{itemize}
\item {\it one-one} method assumes a single state dominates the two-point and three-point
functions. ${\cal A}_0$ and $M_0$ are extracted from a fit to the two-point
function given in Eq.~\ref{eq:two-pt-correlators3} and
$\langle 0 | O_\Gamma | 0 \rangle$ is estimated from
the three-point functions keeping only the first term in Eq.~\ref{eq:three-pt}.

\item {\it Ratio} method also assumes a single state dominates the three-point
function. $\langle 0 | O_\Gamma | 0 \rangle$ are estimated from the
ratio of three-point to two-point functions, which for large $t_\text{sep}$ is
expected to be a constant, the desired matrix element. Some statistical noise may cancel in the ratio as long as the source and sink operators are identical between the two- and three-point functions, but this relies on there being a good signal in both at
separation $t_\text{sep}$.

\item {\it two-two} method: ${\cal A}_0$, ${\cal A}_1$, $M_0$ and $M_1$
  are extracted from a fit to the two-point function. These amplitudes
  and masses are used in a two-parameter fit to the three-point
  function to estimate $\langle 0 | O_\Gamma | 0 \rangle$ and $\langle
  1 | O_\Gamma | 0 \rangle$. In the case of charges where both initial
  and final nucleon operators are at rest, we can assume $\langle 0 |
  O_\Gamma | 1 \rangle$ and $\langle 1 | O_\Gamma | 0 \rangle$ are
  equal, and we analyze only the real part of the three-point
  function. However, in the case of the form factors, the initial and
  final states are not the same and both matrix elements must be
  retained. 

\item The {\it two-sim} method is a simultaneous fit to all $t_\text{sep}$, and the same as the two-two method for extracting ${\cal A}_0$, ${\cal A}_1$, $M_0$ and
  $M_1$. The fit to the three-point function is made to the
  expression in Eq.~\ref{eq:three-pt} using data from all
  investigated values of $t_\text{sep}$ simultaneously. 
   
\item The {\it two-simRR} method --- this simultaneous fit to all $t_\text{sep}$ is the same as the two-sim method but includes the $\langle 1 | O_\Gamma | 1 \rangle$ term.  
 The $\langle 1 | O_\Gamma | 1 \rangle$ term cannot be
  distinguished from the $\langle 0 | O_\Gamma | 0 \rangle$ term if 
  simulations are done at a single $t_\text{sep}$ 
  since it depends only on $t_\text{sep}$, not $t$. This contamination is, however, 
  exponentially suppressed as it is proportional to
  $e^{(M_1-M_0)t_\text{sep}}$.

\end{itemize}
Note that when fitting the form factors, each mass $M_n$ should be replaced by the appropriate energy $E_n$ for the momentum $\vec{p}_i$ used.

In this section, we will briefly demonstrate our analysis method on the isovector charges of $g_{A,S,T}$. 
We will leave the source-sink dependence and various analysis methods used for the isoscalar charges and form factors to the following sections.

\begin{figure*}
\includegraphics[width=0.995\textwidth]{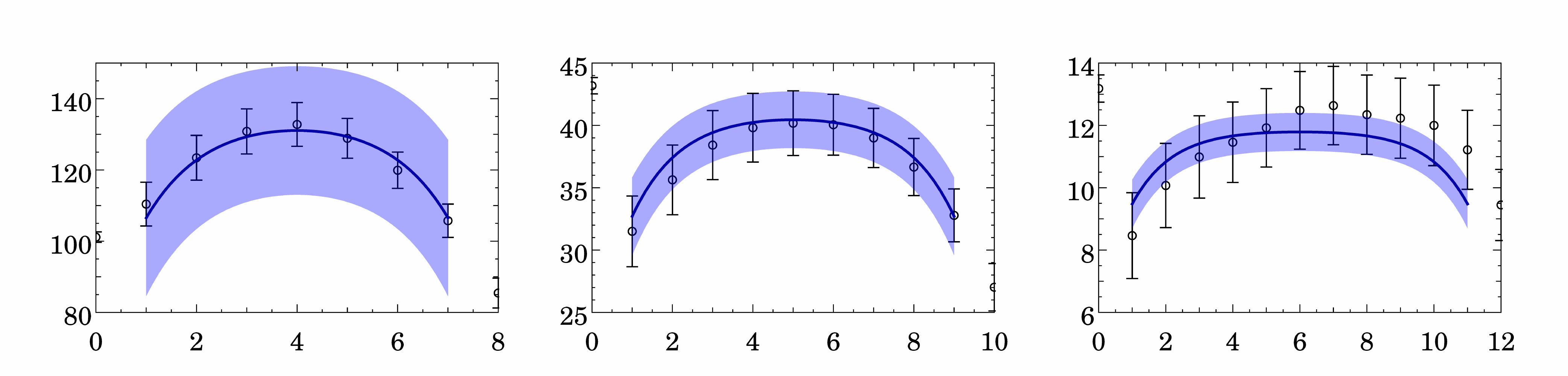}
\includegraphics[width=0.995\textwidth]{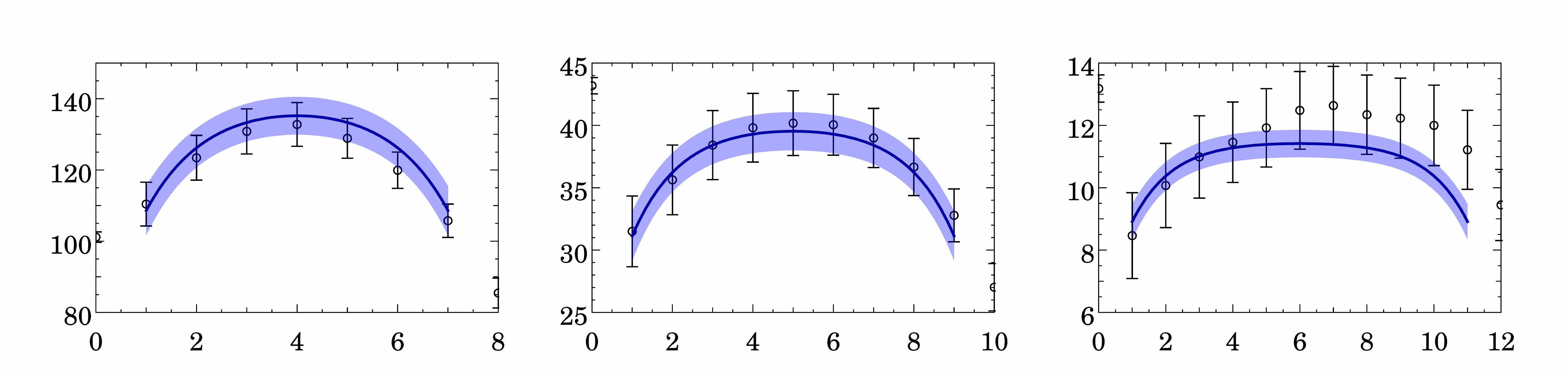}
\caption{ The two-simRR (upper) two-sim (lower) methods fit as a function of time to the unrenormalized $g_S$ data
from the 220-MeV ensemble with insertion on the $d$ quark. The
fits shown are with and without the $\langle 1 |
O_\Gamma | 1 \rangle$ term in Eq.~\ref{eq:three-pt}.
}
\label{fig:gS22fit}
\end{figure*}

We study two versions of the simultaneous fit including excited-state degree of freedom, with and
without the higher-order $\langle 1 | O_\Gamma | 1 \rangle$ term in
Eq.~\ref{eq:three-pt}. Figure~\ref{fig:gS22fit} shows the fit with the worst quality among all our data: 
unrenormalized $g_S$  from the 220-MeV ensemble with (upper) and without (lower) the $\langle 1 | O_\Gamma | 1 \rangle$ contribution. 
Both fits capture the data, and the final fit keeping all the terms is
marginally better. The two-simRR fit is about factor of 2 noisier than 
two-sim one.  We calculate the difference between the two fits
within the jackknife process, and find that for these ensembles the
difference is consistent with zero for all charges at zero momentum with our current statistics. 
To completely illuminate the systematic
error, we will use two-simRR fit for the better determined isovector and 
isoscalar charges. The form factors are noisier than charges, and there
we use only the two-sim fit analysis.

\begin{figure*}
\includegraphics[width=0.995\textwidth]{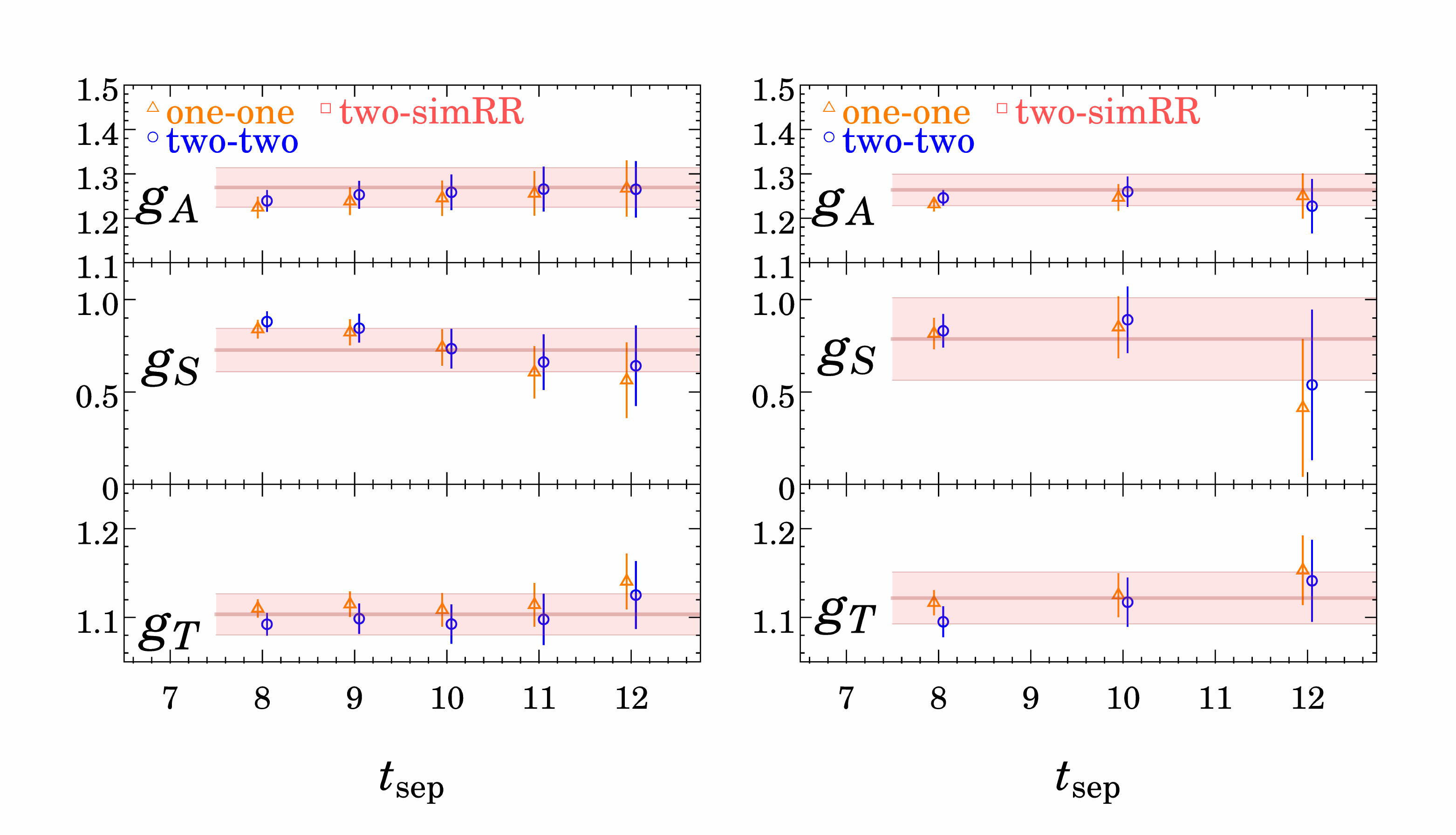}
\caption{ Estimates of the unrenormalized isovector charges $g_{A,S,T}$ as functions of source-sink
  separation ($t_\text{sep}$) with 310-MeV (left) and
  $220$ MeV (right) ensembles at $a \approx 0.12$~fm. Estimates are shown for
  the different fit types described in the text. The band shows the
  results of the two-sim fit to data for all $t_\text{sep}$. }
\label{fig:gSAT}
\end{figure*}

The results of fits for the unrenormalized isovector charges are shown in
Fig.~\ref{fig:gSAT}. Estimates from the two-simRR
method 
are shown by the horizontal bands. 
Based on these two
ensembles of roughly 1000 configurations at $a\approx 0.12$~fm and the
tuned Gaussian-smeared sources used in calculating the quark propagators,
we note the following features for isovector $g_{A,S,T}$:
\begin{itemize}
\item The statistical errors increase by about $40$\% with each unit
increase in $t_\text{sep}$. 

\item Only data for $g_A$ on the 310-MeV ensemble show a small
increase (by about $1 \sigma$) with $t_\text{sep}$ between 8 and 12; $g_S$
shows a decrease of similar magnitude.

\item Based on the trends first seen in the 310-MeV ensemble, we
considered it sufficient to investigate the 220-MeV ensemble
using only $t_\text{sep} \in \{8,10,12\}$.

\item The two-simRR fit estimates of the central values and errors in the isovector charges
are consistent with data from other fits for all values of $t_\text{sep}$ within statistical errors.

\item The errors increase by about 20\% on lowering the light ($u$ and $d$) quark masses
by a factor of two, going from 310- to 220-MeV ensemble.

\item The signal in $g_S$ is the noisiest. Nevertheless, on the 220-MeV ensembles, the error estimate is about $15\%$, reasonably close
to our desired accuracy.
\end{itemize}

Our conclusion, based on the analysis of the $a\approx 0.12$ fm
lattices, is that with a well tuned smeared operator the central
values and error estimates from the two-sim fit agree with those from
the other fits for separation $t_\text{sep}=10$ which corresponds to
$t_\text{sep} \approx 1.2$~fm in physical units for pion mass as light
as 220~MeV in our case. If restricted to simulations at a single
$t_\text{sep}$, we consider the data at $t_\text{sep}=10$ the best
compromise between reducing excited-state contamination and having a
good statistical signal with $O(1000)$ lattices.

Noting the analysis of the bare axial charges presented by the
ETMC~\cite{Alexandrou:2010hf,Dinter:2011sg}, CSSM~\cite{Owen:2012ts}
and LHPC~\cite{Green:2012ej} collaborations, we conclude
that excited-state contamination becomes
comparable to (or smaller than) statistical errors for $t_\text{sep}
\ge 1.2$~fm. These collaborations in their $g_A$ analyses have explored using
summation and variational methods at different values of the lattice
spacing and quark masses and with different number of flavors (see
Table~\ref{tab:chargesothers}). The summation method implemented by the
CLS-Mainz~\cite{Capitani:2012ef,Capitani:2012gj,Brandt:2011sj}
collaboration sums over the full range $t_f$--$t_i$ for multiple $t_\text{sep}$,
including time-slices close to the source and the sink where
excited-state contributions are the largest. Also, their fit ansatz
does not take into account contributions from the transition terms
such as $\langle 0 | O_\Gamma | 1 \rangle $. They conclude that the
summation method gives estimates 1--2$\sigma$ larger than the
``plateau'' (ratio) method with $t_\text{sep} \approx 1$~fm. However, examining their data in
detail at each $t_\text{sep}$, the various estimates are consistent within errorbars.

\section{Nonperturbative Renormalization in RI-sMOM Scheme}
\label{sec:Zfac}

This section describes the lattice calculation of the renormalization
constants $Z_{A,S,T}$ in the RI-sMOM scheme
(regularization-independent symmetric
momentum-subtraction)~\cite{Martinelli:1994ty,Sturm:2009kb} and their
conversion to $\overline{\text{MS}}$ scheme at $2$~GeV. For this
calculation, the HISQ configurations are fixed to Landau gauge after
hypercubic (HYP) smearing and the clover propagators $S(0,x)$ are calculated using
point sources. From these propagators, we extract the wavefunction
renormalization constant $Z_\psi$ and calculate the truncated three-point
correlators as functions of renormalization scale $\mu$ to
estimate the three $Z_{A,S,T}$.

The clover action we use is improved to $O(a)$ only in tadpole-improved
perturbation theory. Our renormalized operators, defined as
$O_\Gamma^\text{R} = Z_\Gamma O_\Gamma$, also do not include any $O(a)$
improvements. The errors in the results, therefore, start at $O(a)$. 
The three-point function we calculate is defined by
\begin{equation}
\Lambda_\Gamma(x, 0, y) = \psi(x) O_\Gamma(0) \overline{\psi}(y) = S(x, 0) \Gamma S(0, y) ,
\end{equation}
with $\Gamma$ representing the Dirac matrices $I$ (scalar),
$\gamma_\mu\gamma_5$ (axial-vector) and $\sigma_{\mu \nu}$ (tensor).
In momentum space, this three-point function is
\begin{equation}
\Lambda_\Gamma(p_i, p_f) = S(p_i) \Gamma (\gamma_5 S^\dagger(p_f) \gamma_5)  ,
\end{equation}
where $S(p)$ is the Fourier transform of $S(x,0)$, and we have applied
$\gamma_5$-Hermiticity to the right quark leg. From this we
construct the amputated three-point correlator $\Lambda^\text{A}_\Gamma(p_i,p_f)$
\begin{multline}
\Lambda^\text{A}_\Gamma(p_i, p_f) =\\
 S(p_i)^{-1} S(p_i) \, \Gamma \, (\gamma_5 S^\dagger(p_f) \gamma_5) (\gamma_5 S^\dagger(p_f) \gamma_5)^{-1},
\end{multline}
and the projected amputated three-point function
\begin{equation}
\label{eq:pa-correlator}
\Lambda^\text{PA}_\Gamma(p_i, p_f) = \frac{1}{12} \Tr \Big(P_\Gamma \Lambda^\text{A}_\Gamma(p_i, p_f) \Big), 
\end{equation}
where the projector $P_\Gamma$ for the RI-sMOM scheme is 
$I$ (scalar), $ (q_\mu/{q^2}) \gamma_5 \slashed{q}$ (axial-vector) and 
$(i/12)\gamma_{[\mu} \gamma_{\nu]}$ (tensor).
In the RI-sMOM scheme, the allowed momenta satisfy the relations 
\begin{equation}
\label{eq:mom}
p_f^2 = p_i^2 = q^2 , \qquad q=p_f-p_i \neq 0 .
\end{equation}

The renormalized projected amputated three-point function is defined as 
\begin{multline}
\left.\Lambda^\text{R}_\Gamma(p_i, p_f) \right|_{p_i^2 = p_f^2 = q^2} = \\
\left.\left(Z_\psi^{-1}Z_\Gamma~\Lambda^\text{PA}_\Gamma(p_i, p_f) \right)\right|_{p_i^2 = p_f^2 = q^2} ,
\end{multline}
where $Z_\Gamma$ is the operator renormalization
constant. In the RI-sMOM scheme, this is set equal to one, its
tree-level value, for all tensor structures. This condition fixes the
value $Z_\psi^{-1} Z_\Gamma$ at the subtraction point. Similarly, 
the wavefunction renormalization constant $Z_\psi$ is defined by
\begin{equation}
\left.(Z_\psi)^{-1} \frac{i}{12} \Tr \left(\frac{\slashed{p}  S(p)^{-1}  }{p^2}\right)\right|_{p^2 = q^2} = 1 .
\end{equation}

Having extracted the renormalization constants in the RI-sMOM scheme
at scale $\mu=\sqrt{q^2}$, we convert them to the $\overline{\text{MS}}$ scheme
at $\mu = 2$~GeV using 
the one-loop conversion factors in Landau gauge~\cite{Sturm:2009kb, Gracey:2011fb}:
\begin{align}
C_\psi &= 1, \\
C_A &= 1, \\
C_S &= 1 + \frac{\alpha_s(\mu)}{4 \pi} \frac{4}{3} \left[4 + \frac{2}{3} \pi^2 - 10.0956 \right], \\
C_T &= 1 - \frac{\alpha_s(\mu)}{4 \pi} \frac{4}{3} \left[\frac{1}{3} \left(4 + \frac{2}{3}\pi^2 - 10.0956 \right) \right],
\end{align}
where $\alpha_s(\mu)$ in this horizontal matching can, to
$O(\alpha_s^2)$, be taken to be the coupling in any scheme, RI-sMOM or
$\overline{\text{MS}}$ scheme or from the plaquette using the BLM
procedure~\cite{Brodsky:1982gc}. We use the coupling
$\alpha_{\overline{\text{MS}}}$.  Note that the above conversion factors
are computed in the chiral limit, and possible $O(ma)$ corrections
are ignored.

These $Z$, now in $\overline{\text{MS}}$ scheme defined at scale $\mu'=\sqrt{q^2}$, are then run to $\mu=2$~GeV using
\begin{equation}
Z_\Gamma(\mu) = \frac{E_\Gamma(\frac{\alpha_s(\mu)}{\pi})}{E_\Gamma(\frac{\alpha_s(\mu')}{\pi})}Z_\Gamma(\mu'),
\end{equation}
where the evolution function ${E}_\Gamma(\alpha_s(\mu)/\pi)$ at two-loop is~\cite{Chetyrkin:1997dh}
\begin{multline}
E_\Gamma\left(\frac{\alpha_s(\mu)}{\pi}\right) = \\
\left(\frac{\alpha_s(\mu)}{\pi}\right)^{\frac{\gamma^0_\Gamma}{\beta_0}}\left[1 + \left(\frac{\gamma^1_\Gamma}{\beta_0} - \frac{\beta_1}{\beta_0}\frac{\gamma^0_\Gamma}{\beta_0}\right)\frac{\alpha_s(\mu)}{\pi}\right],
\end{multline}
$\beta_0$ and $\beta_1$ are the universal coefficients of the $\beta$-function,
\begin{align}
\beta_0 &= \frac{1}{12}(11C_A - 4T_F n_f), \\
\beta_1 &= \frac{1}{24}(17C_AC_A - 10C_A T_F n_f - 6C_F T_F n_f),
\end{align}
with $C_A=3$, $C_F = 4/3$ and $T_F=1/2$; $\gamma^0_\Gamma$ and $\gamma^1_\Gamma$ are the first two coefficients in the anomalous dimension of the operators in the $\overline{\text{MS}}$ scheme,
\begin{align}
\gamma^0_A& = 0, \qquad \gamma^1_A = 0, \\
\gamma^0_S &= -\frac{3}{4}C_F, \nonumber\\
 \gamma^1_S &= -\frac{1}{16}\left(\frac{3}{2}C_F^2 + \frac{97}{6}C_FC_A -\frac{10}{3}C_FT_Fn_f\right), \\
\gamma^0_T &=  \frac{1}{4}C_F, \nonumber\\
 \gamma^1_T &= \frac{1}{16}\left(-\frac{19}{2}C_F^2 + \frac{257}{18}C_FC_A - \frac{26}{9}C_FT_Fn_f\right),
\end{align}
and $\alpha_s(\mu)$ at two-loop has the following expression
\begin{equation}\label{eq:alpha-2loop-running}
\alpha_s(\mu) = \frac{\alpha_s(\mu')}{v(\mu)} \left[1 - \frac{\beta_1}{\beta_0} \frac{\alpha_s(\mu')}{4\pi} \frac{\ln v(\mu)}{v(\mu)}\right],
\end{equation}
where
\begin{equation}
v(\mu) = 1 - \beta_0 \frac{\alpha_s(\mu')}{2\pi} \ln \left(\frac{\mu'}{\mu}\right).
\end{equation}

\begin{figure}
\includegraphics[width=0.45\textwidth]{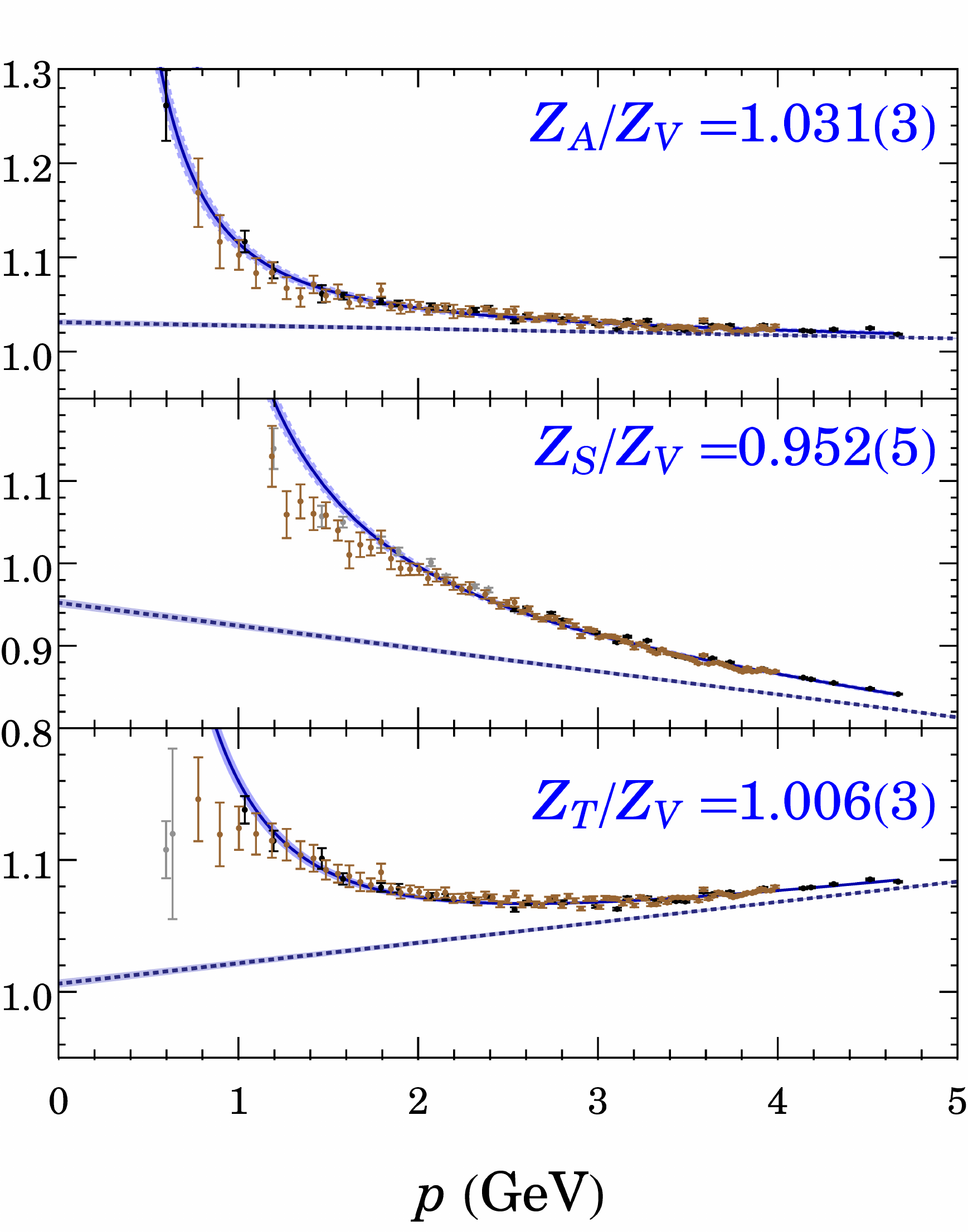}
\caption{The data for $Z_{A,S,T}/Z_V$ in the $\overline{\text{MS}}$ scheme 
at $2$~GeV and fits using the ansatz $ c/q^2 + Z + d_1 q $. 
The data for the $M_\pi=310$ ($220$) ensemble are shown by black (brown) symbols. 
In each case the straight line is the plot of $Z + d_1 q $ where $Z$ is 
the listed value of $Z_\Gamma/Z_V$. }
\label{fig:Z310}
\end{figure}

We have analyzed 101 (60) lattices on the $M_\pi=310$~MeV (220~MeV) ensemble and 
plot the results for the ratios $Z_\Gamma/Z_V$ in Fig.~\ref{fig:Z310}. In 
extracting the final estimates, we incorporated the following observations:
\begin{itemize}
\item
The data for the ratios $Z_\Gamma/Z_V$ are more smooth in all cases, and the 
fits are more stable. Some sources of systematic uncertainty cancel in the ratios.
\item
We find no discernible difference as a function of the pion mass, and
the fits to the two data sets are indistinguishable as shown in
Fig.~\ref{fig:Z310} where both sets are plotted together. We, therefore, neglect 
possible mass dependence and make a single fit to the combined data. 
\item
The estimates of $Z_\Gamma$ are sensitive to the different possible
combinations of momentum components $p_\mu$ for a given value of
$p^2=q^2$. These differences are indicative of lattice discretization
effects due to reduction of the continuum Lorentz group to the
hypercubic group on the lattice. Since these effects are smallest in
momenta with symmetric components, for example the combination
$\{1,1,1,1\}$ versus $\{2,0,0,0\}$, we choose the most symmetric
combinations for our final analysis, shown in Fig.~\ref{fig:Z310}. That
is, for a given $p^2$, we choose momenta that minimize
$p_{1,2}^{[4]}/(p^2)^2$ with $p^{[4]} = \sum_\mu p^4_\mu$.
\end{itemize}

At weak enough coupling the final results for the renormalization
constants in the $\overline{\text{MS}}$ scheme at $2$~GeV should be
independent of the $q^2$ value (within a window $\Lambda_{\rm
QCD} \ll \sqrt{q^2} \ll \pi/a$) selected to define them in the RI-sMOM
scheme. This requires that the errors due to lattice discretization
and the use of truncated perturbation theory to convert from RI-sMOM
to $\overline{\text{MS}}$ scheme are negligible. The data for $Z_S/Z_V$,
however, show very significant $q^2$ dependence. This calls into
question whether on $0.12$-fm lattices there exists a window
in which the
nonperturbative effects at low $q^2$ and lattice-discretization effects at high
$q^2$ are both small.

Recent analyses have shown that smeared lattices alter the window
in which lattice data are consistent
with perturbation theory~\cite{Arthur:2013bqa}.
HYP smearing smooths
the gauge fields in a $2^4$ hypercube, so gluons with momentum
of order $1/a$ are suppressed,
modifying the high-$q^2$ behavior of $Z_\Gamma$. Since our
data for $Z_S/Z_V$ shows large $q^2$ dependence, we investigate two
prescriptions, each defined such that the renormalized charges,
$Z_\Gamma/Z_V \times g_\Gamma^\text{bare}/g_V^\text{bare}$, have a well-defined continuum
limit. In the first case we pick the value of $Z_\Gamma/Z_V$ at a
fixed physical value, $q^2 = 5\text{ GeV}^2$. The error on this
estimate is taken to be half of the total variation in the range $4 < q^2 <
6\text{ GeV}^2$. In the second case we
assume an ansatz for the lattice artifacts and fit the $q^2$ dependence of the
data~\cite{Arthur:2013bqa}. We try several variations of $ c/q^2 +
Z + d_1 q + d_2 q^2$ and of the fit range. We do not include
dependence on the pion mass, since the data do not show any
significant difference between the two ensembles. We find that the
ansatz $ c/q^2 + Z + d_1 q $ is a good fit to all the data for $q^2 >
1\text{ GeV}^2$, as shown in Fig.~\ref{fig:Z310}. We take the estimates 
of $Z$ from these fits as our central values. In addition to fits to the
ratios $Z_\Gamma/Z_V$, we also constructed the ratios, within a
jackknife process, from fits to individual $Z_\Gamma$.

\begin{table}
\begin{tabular}{|cc|ccc|}
\hline
$Z_\psi$     & $Z_V$   & $Z_A/Z_V$   & $Z_S/Z_V$   & $Z_T/Z_V$     \\
\hline                   
0.98(1)      & 0.89(3) & 1.03(1)     & 0.95(3)     & 1.01(2)   \\
\hline                                      
\end{tabular}
\caption{The results for $Z_\psi$, $Z_V$, and the three ratios of renormalization constants 
$Z_{A, S, T}/Z_V$ in the $\overline{\text{MS}}$ scheme at 2~GeV. The
lattice calculation is done in the RI-sMOM scheme. The error estimates
cover the spread in values from different methods used as discussed in
the text. We do not find significant variation with the pion mass and
quote a single result that is used for both ensembles.}
\label{tab:resultsZ}
\end{table}

It turns out that the three estimates are consistent. This is easy to
see, from data shown in Fig.~\ref{fig:Z310}, for $Z_A/Z_V$ and
$Z_T/Z_V$ since the $q^2$ dependence above $q^2 = 4\text{ GeV}^2$ is
small.

To assign an overall error to $Z_\Gamma/Z_V$ we took into account 
the different estimates. Our final error estimate covers (i) 
the variation with $q^2$ between $4 \le q^2 \le 6\text{ GeV}^2$ in method one, (ii) 
the variation with the fit ansatz in method two and (iii) the difference in
the three estimates.  Note that this conservative error estimate,
given in Table~\ref{tab:resultsZ}, is much larger than the error from
the fits as shown in Fig.~\ref{fig:Z310}. It also captures the
spread in data corresponding to points with different breaking of rotational
symmetry, {\it i.e.} data with larger $p_{1,2}^{[4]}/(p^2)^2$ values.

Our final estimate for the renormalized charges is obtained by
multiplying the ratios $g_\Gamma^\text{bare}/g_V^\text{bare}$ with the
corresponding ratios $Z_\Gamma/Z_V$ since the vector Ward identity
implies $Z_V g_V^\text{bare} = 1$.

\section{Nucleon Isovector Charges}
\label{sec:charges}

We present results for the three unrenormalized charges from two-simRR fit in
Table~\ref{tab:resultsG} and the final renormalized values in
Table~\ref{tab:gresultsfinal}. To facilitate comparison with previous 
work with improved actions, we give a compilation of lattice parameters used by
other collaborations in Table~\ref{tab:chargesothers}, and
selected results
are shown in Fig.~\ref{fig:gA}.

We employ two strategies to extract renormalized charges $g_{A,S,T}$,
and their difference is used as an estimate of systematic errors. In
the first method, we extract, under two separate jackknife analyses
due to the different numbers of configurations analyzed, the
unrenormalized charges $g_{A,S,T}^\text{bare}$ and the renormalization
constants $Z_{A,S,T}$ in the $\overline{\text{MS}}$ scheme at
$2$~GeV. These are multiplied together with relative errors added in
quadrature. In the second method, we extract the ratios
$g_{A,S,T}^\text{bare}/g_V^\text{bare}$ and $Z_{A,S,T}/Z_V$ and use the
fact the $Z_V g_V^\text{bare} = 1$. Our data, however, give $Z_V g_V = 0.95(3)$
and $0.96(4)$ for the 310- and 220-MeV ensembles, respectively; this
leads to a difference of up to $0.04$ between the two estimates.

\begin{table*}
\small\setlength{\tabcolsep}{1pt}
\hfuzz=3pt
  \begin{tabular}{|lccccccc|}
\hline 
Collaboration                   & Action         & $N_f$ & $M_\pi$ (MeV) & $L$ (fm)     & $(M_\pi L)_\text{min}$ & $a$ (fm)                & Charges Calculated  \\
\hline\hline 
QCDSF\cite{Khan:2006de}         & clover         & 2     & 595--1000     & 1.0--2.0     & $4.6$                 & 0.07--0.116             &  $g_A$ \\
ETMC\cite{Alexandrou:2010hf}    & twisted Wilson & 2     & 260--470      & \{2.1, 2.8\} & $3.3$                 & $\{0.056,0.070,0.089\}$ &  $g_A$        \\
QCDSF~\cite{Pleiter:2011gw}     & clover         & 2     & 170--270      & 2.1--3.0     & $2.6$                 & 0.08--0.116             &  $g_A$, $g_T$ \\
CLS-Mainz\cite{Capitani:2012ef,Capitani:2012gj,Brandt:2011sj} & clover         & 2     & 277--649      & 2.0--3.0     & $4.0$                 & $\{0.05,0.063,0.079\}$    &   $g_A$ \\
QCDSF\cite{Horsley:2013ayv}     & clover         & 2     & 157--1600     & 0.86--3.42   & $2.64$                & 0.06--0.075             &  $g_A$  \\
RBC\cite{Lin:2008uz}            & DWF            & 2     & 490--695      & 1.9          & $4.75$                & 0.117                   &  $g_A$, $g_T$ \\ 
\hline 
RBC/UKQCD\cite{Yamazaki:2008py,Aoki:2010xg}           & DWF                & 2+1 & 330--670 & $\{1.8,2.7\}$     & $3.8$   & 0.114               & $g_A$, $g_T$  \\
LHPC\cite{Edwards:2005ym,Edwards:2006qx,Bratt:2010jn} & DWF on staggered   & 2+1 & 290--870 & $\{2.5,2.7\}$     & $3.68$  & 0.1224              & $g_A$, $g_T$ \\
QCDSF\cite{Gockeler:2011ze}                          & clover             & 2+1 & 350--480 & 1.87              & $3.37$  & 0.078               & $g_A$  \\
HSC\cite{Lin:2011sa}                                  & anisotropic clover & 2+1 & 450--840 & 2.0               & $4.57$  & 0.125 ($a_t=0.036$) & $g_A$  \\
CSSM\cite{Owen:2012ts}               & clover         & 2+1   & 290      & 2.9            & 4.26                                      & 0.091           & $g_A$  \\
LHPC\cite{Green:2012ej,Green:2012ud} & clover         & 2+1   & 149--357 &  $\{2.8,5.6\}$ & 3.57   & \{0.116, 0.09\} & $g_A$,  $g_S$, $g_T$ \\
\hline 
ETMC\cite{Dinter:2011sg,Alexandrou:2013joa} & twisted Wilson & 2+1+1 & 354--465      & 2.5--2.9            & 3.35                                      & 0.066--0.086         & $g_A$  \\
PNDME (this work)                    & clover on HISQ & 2+1+1 & 220--310 & \{2.88,3.84\}  & 4.28                                      & 0.12            & $g_A$,  $g_S$,  $g_T$ \\
\hline 
\end{tabular}
\caption{A summary of the lattice parameters used by various collaborations in the calculations of charges $g_{A,S,T}$.}
\label{tab:chargesothers}
\end{table*}

\begin{table*}
\begin{tabular}{|c|cccc|ccc|}
\hline
              & $g_V^\text{bare}$ & $g_A^\text{bare}$ & $g_S^\text{bare}$ & $g_T^\text{bare}$  & $g_A^\text{bare}/g_V^\text{bare}$ & $g_S^\text{bare}/g_V^\text{bare}$ & $g_T^\text{bare}/g_V^\text{bare}$  \\\hline
310-MeV & $1.068(13)$  & $1.269(45)$  & $0.73(12)$   & $1.104(23)$   & $1.189(45)$  & $0.68(11)$   & $1.034(22)$   \\
\hline
220-MeV & $1.081(16)$  & $1.264(35)$  & $0.79(22)$   & $1.122(29)$   & $1.169(35)$  & $0.73(20)$   & $1.038(30)$   \\
\hline
\end{tabular}
\caption{The final results based on the two-simRR method (including $\langle 1 | O_\Gamma | 1 \rangle$ term in Eq.~\ref{eq:three-pt}) for the three unrenormalized charges 
$g_{A,S,T}^\text{bare}$ and their ratios to $g_V^\text{bare}$.
The errors quoted are statistical from an overall single-elimination jackknife procedure.}
\label{tab:resultsG}
\end{table*}

\begin{table}
\begin{tabular}{|c|ccc|}
\hline
                & $g_A$                 & $g_S$                & $g_T$              \\\hline
310-MeV         & $1.226(48)$           & $0.65(10)$           & $1.040(30)$        \\\hline
220-MeV         & $1.205(38)$           & $0.69(19)$           & $1.044(36)$        \\\hline\hline
Extrapolation   & $1.193(68)$           & $0.72(32)$           & $1.047(61)$         \\\hline
\end{tabular}
\caption{The final results for the three charges obtained by combining the ratios  
$(Z_\Gamma/Z_V)(g_\Gamma^\text{bare}/g_V^\text{bare})$ and using $Z_V g_V^\text{bare} = 1$ as discussed in the text.
The error quoted is obtained by combining the statistical
and systematic errors in the ratios of $Z$'s and $g$'s in quadrature under
the assumption that they are independent. 
The last row gives estimates extrapolated to the physical
pion mass $M_\pi=140$~MeV using a fit linear in $M_\pi^2$.
}
\label{tab:gresultsfinal}
\end{table}

\subsection{Axial Charge $g_A$}
\label{subsec:gA}

\begin{figure*}
\includegraphics[width=0.32\textwidth]{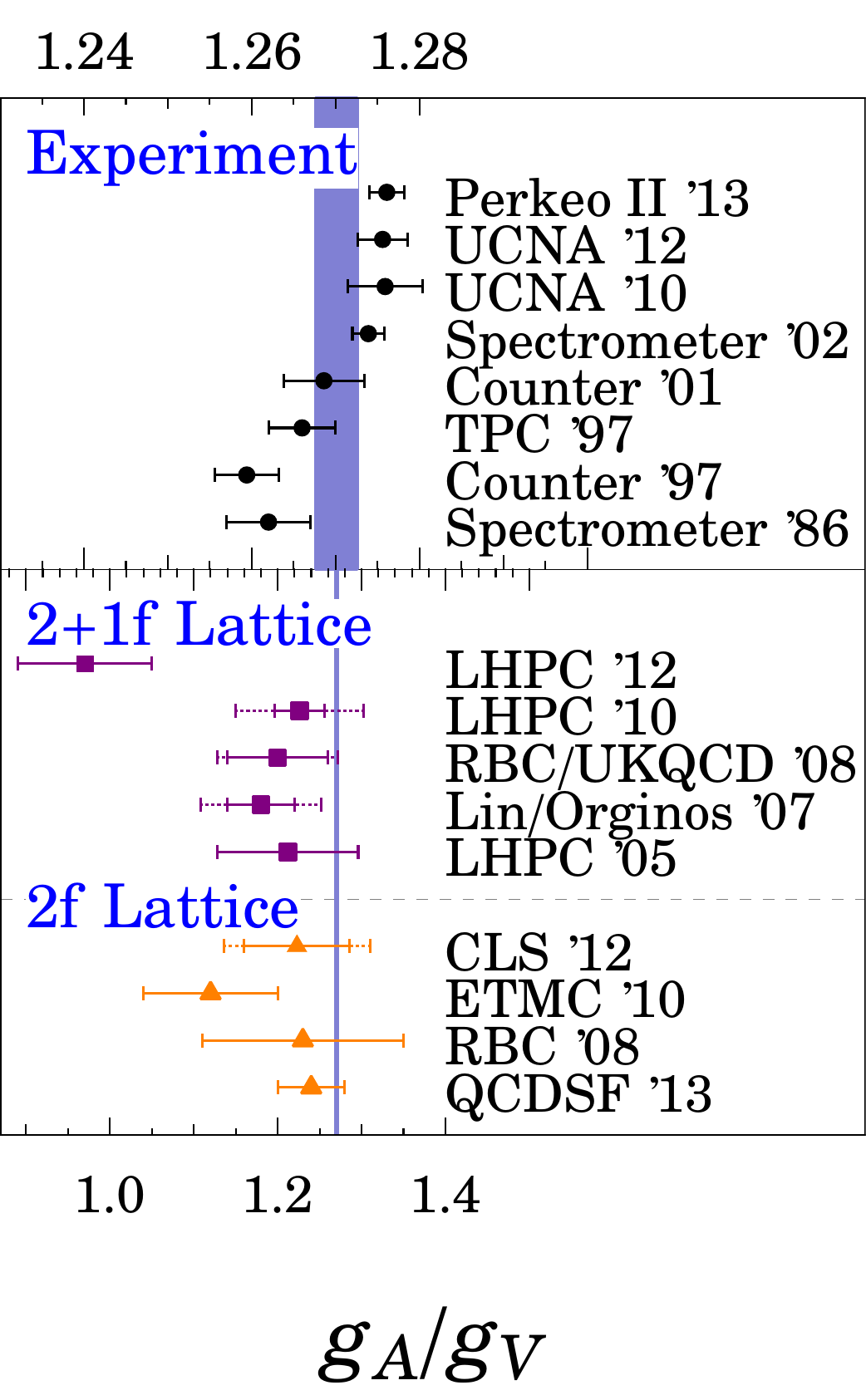}
\includegraphics[width=0.64\textwidth]{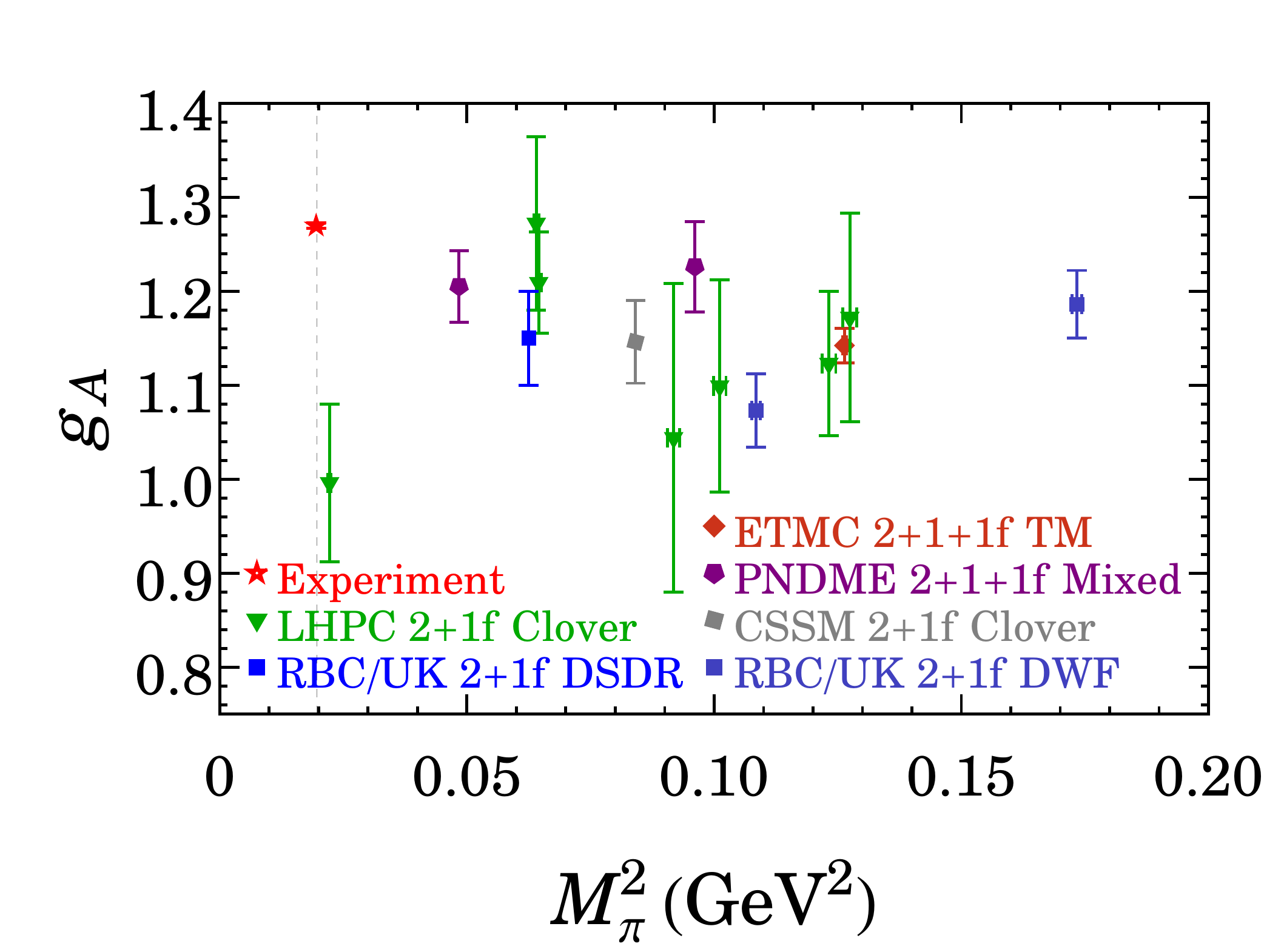}
\caption{ (Left) Collected experimental values used in PDG~2012
  average (the band) and the latest UCNA (2012) measurements on $g_A$;
  there has been a slow increase in $g_A/g_V$ over the past 15
  years. The lower panel shows $g_A$ values after extrapolating
  to the physical pion mass collected from dynamical 2+1-flavor and 2-flavor lattice
  calculations using $O(a)$-improved
  fermions~\cite{Khan:2006de,Lin:2008uz,Alexandrou:2010hf,Capitani:2012gj,Edwards:2005ym,Lin:2007ap,Yamazaki:2008py,Bratt:2010jn,Green:2012ud}.
  Note the change in scale between the experimental and theory plots. Most of the errorbars here are
  statistical only. In data from the few calculations that also quote systematic errors, we add these
  to the statistical ones as outer errorbar bands, marked with
  dashed lines. (right) Calculations of $g_A$
  using at least 2+1 flavors $O(a)$-improved dynamical fermions, plotted as a function of
  $M_\pi^2$, with $m_\pi L > 4$ to avoid systematics due to small spatial extent.}
\label{fig:gA}
\end{figure*}

The best lattice-QCD calculations, involving multiple lattice spacings
(including continuum extrapolation of estimates) and high statistics,
yield estimates with about 5\% statistical error and are a few standard
deviations lower than the experimental values. As shown on the
right-hand side of Fig.~\ref{fig:gA}, most lattice estimates lie
between $1.05$ and $1.20$, which is 5--15\% below the experimental values.

Most collaborations, such as
ETMC~\cite{Alexandrou:2010hf,Dinter:2011sg}, CSSM~\cite{Owen:2012ts},
CLS-Mainz~\cite{Capitani:2012ef,Capitani:2012gj,Brandt:2011sj},
and LHPC~\cite{Green:2012ej}, find that estimates of central value on $g_A$ increase with
nucleon source-sink separation $t_\text{sep}$ in three-point function and that the statistical errors also grow.
Our two-simRR fit result
agrees with estimates from the larger values of $t_\text{sep} \ge 10$ and
after multiplication by $Z_A$ gives $1.193(68)$ as shown in
Table~\ref{tab:gresultsfinal}. This estimate, without continuum extrapolation, is
about $1 \sigma$ below the experimental value.

It has previously been shown that estimates of $g_A$ can be
underestimated due to insufficiently large spatial volumes (see
Refs.~\cite{Yamazaki:2009zq,Horsley:2013ayv} for example), especially
those with $M_\pi L < 4$. Finite-volume corrections based on HBXPT
\cite{Alexandrou:2013joa,Horsley:2013ayv} or a simple parametrization
formula~\cite{Yamazaki:2009zq} are often used to correct this
systematics.  However, a recent global survey on lattice
$g_A$~\cite{Lin:2012ev} suggested that there might exist ambiguities
in the HBXPT correction and that larger volume $M_\pi L \approx 6$
might be needed to reproduce the experimental value of $g_A$. Thus, a
more detailed finite-volume study with pion masses below 250~MeV is
needed to understand this systematic better.

In their most recent work~\cite{Horsley:2013ayv}, QCDSF collaboration
find $g_A=1.24(4)$ based on a new data point at the physical
pion mass. Data at and above their next lightest
$M_\pi=253$~MeV ensemble lie in the range $1.05 \le g_A \le 1.10$. 
Their chiral fit, consequently, suggests that $g_A$ increases 
significantly between 140 and 250~MeV. 
Recent ETMC results~\cite{Alexandrou:2013joa} do not
show a significant increase between 210 and 250~MeV. To clarify the
chiral behavior, therefore, requires data below 210~MeV.

To summarize, current data suggest that to obtain a precise value of
$g_A$ will require simulations close to the physical light-quark
masses, large lattices, high statistical precision and a careful study
of excited-state contamination.

\subsection{Scalar and Tensor Charges $g_S$ and $g_T$}
\label{subsec:gSgT}

\begin{figure}
\includegraphics[width=.48\textwidth]{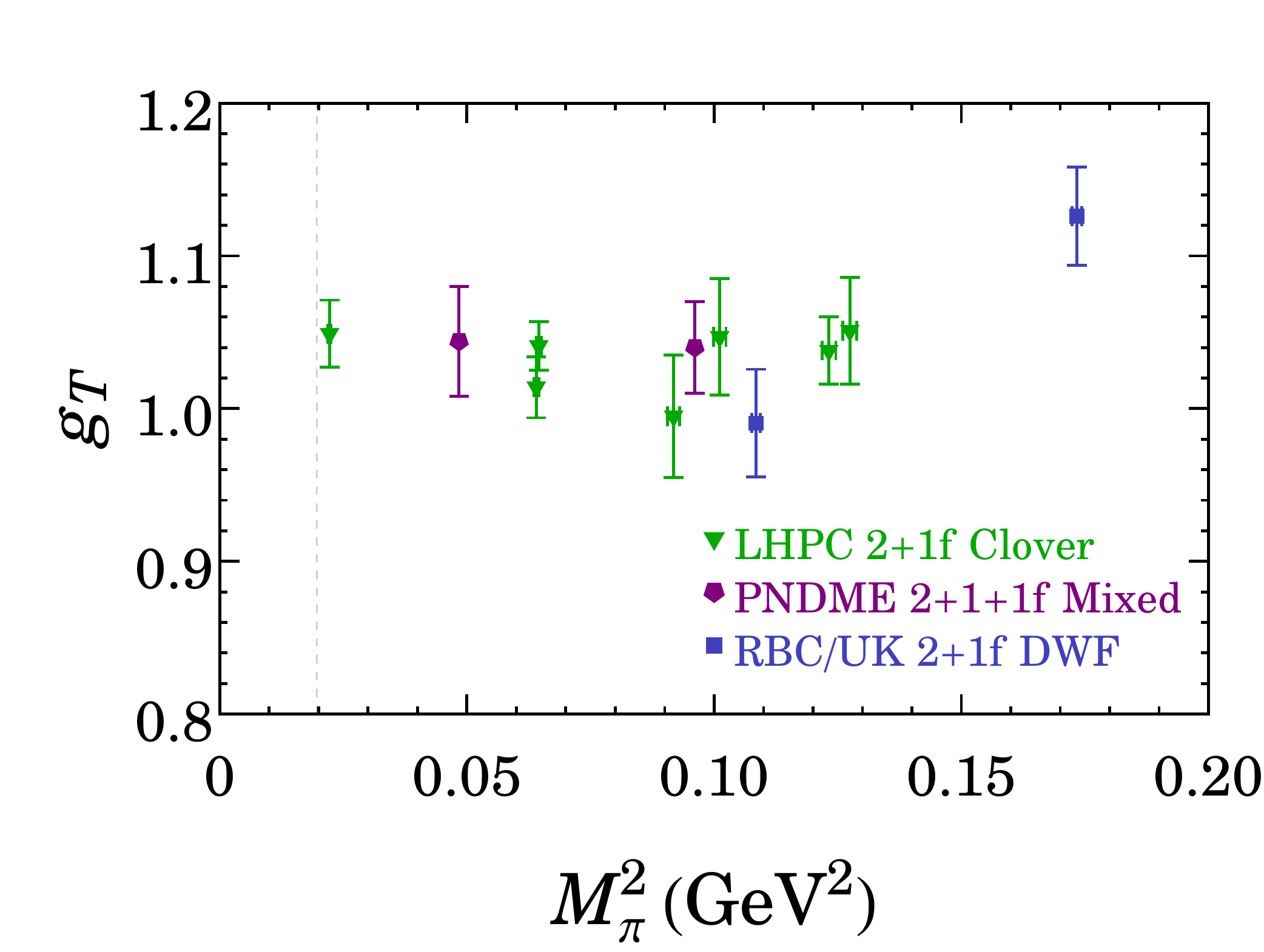}
\includegraphics[width=.48\textwidth]{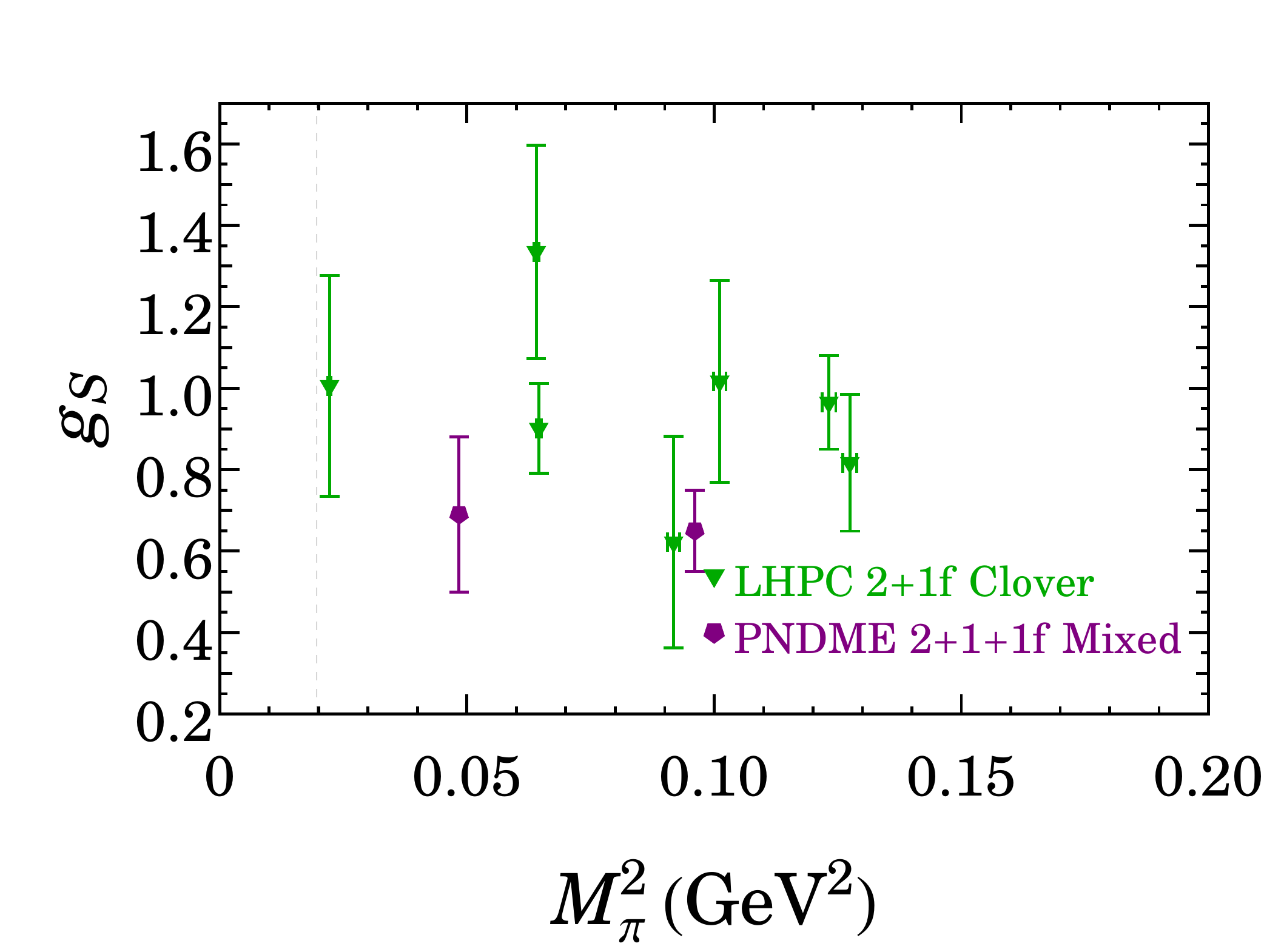}
\caption{
Global analysis of all $N_f=2+1(+1)$ lattice calculations of $g_T$ (above) and $g_S$ (below) with $m_\pi L > 4$ to avoid systematics due to small spatial extent. The dashed line indicates the location of the physical pion mass.}
\label{fig:gTS}
\end{figure}

Our final estimates, given in Table~\ref{tab:gresultsfinal} and shown in Fig.~\ref{fig:gTS}, are
$g_S = 0.72(32)$ and $g_T = 1.047(61)$.
LHPC has recently published lattice calculations giving
$g_S=1.08(28)(16)$ and $g_T=1.038(11)(12)$~\cite{Green:2012ej}. Their main
result exploits a single ensemble created with tree-level
clover-improved Wilson fermions on $48^4$ lattices at a very similar
value of the lattice spacing $a\approx 0.116$~fm but at $M_\pi\approx 150$~MeV.
Their result~\cite{Green:2012ud} for $g_A$ on the same lattices is near 1 (1.00(8)),
which, combined with current experimental values for the neutron lifetime,
implies $V_{ud} > 1$. It seems evident that not all systematic
uncertainties are under control for this ensemble.
One should also note that both calculations lack continuum
extrapolation. Our future update will include $0.09$-fm and $0.06$-fm
ensembles and thus reduce the systematic uncertainty due to the continuum
extrapolation.  The phenomenological implications of the estimates of 
$g_S$ and $g_T$ are given in section~\ref{sec:end}.

\section{Nucleon Connected Isoscalar Charges}
\label{sec:isoscalar}

In this section we summarize results for the connected diagrams
contributing to the isoscalar charges $g^s_{A,S,T}$. One motivation
for their study is that $g^s_T$ probes novel contributions to the quark electric dipole
moment inside the nucleon~\cite{Bhattacharya:2012bf}, as discussed below.

The neutron electric dipole moment (nEDM) $d_n$ is a measure of the distribution of
positive and negative charge inside the neutron. To generate a nEDM,
the theory must include processes that violate CP-symmetry. There are
two sources of CP-violation in the Standard Model: the phase in the
CKM matrix and the $\Theta$-term in the Lagrangian. The CKM phase
gives rise to a nEDM $d_n \sim 10^{-32}$~$e\cdot\mbox{cm}$~\cite{Dar:2000tn} that is too
small to account for the observed baryon asymmetry of the
universe. The current upper limit $\Theta < 10^{-10}$, an
unnaturally small number, is obtained from the current experimental
limit $d_n < 2.9 \times 10^{-26}$~$e\cdot\mbox{cm}$~\cite{Baker:2006ts}. Possible new
interactions at the TeV scale (supersymmetry, left-right models,
extra dimensions) are a rich source of additional CP violation that
could give rise to a large nEDM in the range $10^{-28}$--$10^{-26}$~$e\cdot\mbox{cm}$, enough to
explain baryogenesis. This is an exciting scenario, since the next
generation of EDM experiments are targeting $10^{-27}$~$e\cdot\mbox{cm}$.

Independent of the details of the candidate theories at the TeV scale,
in the effective field theory language, there are two CP-violating
operators at dimension five that give rise to CP-violating
interactions of the electric field with the neutron. These are the
quark EDM (qEDM) and quark chromoelectric dipole moment (CEDM) operators~\cite{Pospelov:2005pr,Bhattacharya:2012bf}
\begin{multline}
        {i\,e} \frac{v_H}{\Lambda_\text{BSM}^2} \sum_{q=u,d} d_q^\gamma {\bar q} \sigma_{\mu\nu}\gamma_5 F^{\mu\nu} q + \\
        {i\,g_3} \frac{v_H}{\Lambda_\text{BSM}^2} \sum_{q=u,d} d_q^G {\bar q} \sigma_{\mu\nu}\gamma_5 \lambda^A G^{\mu\nu\, A} q .
\label{eq:eft}
\end{multline}
Here, $F^{\mu\nu}$ is the electromagnetic field, $G^{\mu\nu}$ is the
gluon field and $e$ and $g_3$ are their respective couplings. The
couplings, $\{d_u^{\gamma,G}, d_d^{\gamma,G}\}$, encapsulate the
interaction of the quarks with the photon and gluon, and $v_H = 246$~GeV
is the vacuum expectation value of the Higgs field. The matrix
elements of these operators are very poorly known~\cite{Engel:2013lsa}
and are needed in order to use future measurements of nEDM to tighten
constraints on the allowed parameter space of BSM theories.

The matrix element of the qEDM operator is an extension of the
lattice-QCD calculation of $g_T$; one needs to calculate 
terms of the form
\begin{equation}
\left.\left\langle n \middle| J^\text{EM}_\mu \middle| n \right\rangle\right|_{\cancel{\mbox{CP}}}^\text{qEDM}
= p^\nu \sigma_{\mu\nu} d_n
= p^\nu \sum_q d_q^\gamma \left\langle n \left| \bar q \sigma_{\mu\nu} q \right| n  \right\rangle ,
\label{eq:qEDM}
\end{equation}
which can be expressed in terms of the isoscalar and isovector
tensor charges of the neutron. We have already discussed the
calculation of the isovector tensor charge $g_T$, and present first
results for the connected part of isoscalar tensor charge. The
remaining disconnected part is beyond the scope of this study.

\begin{figure*}
\includegraphics[width=0.90\textwidth]{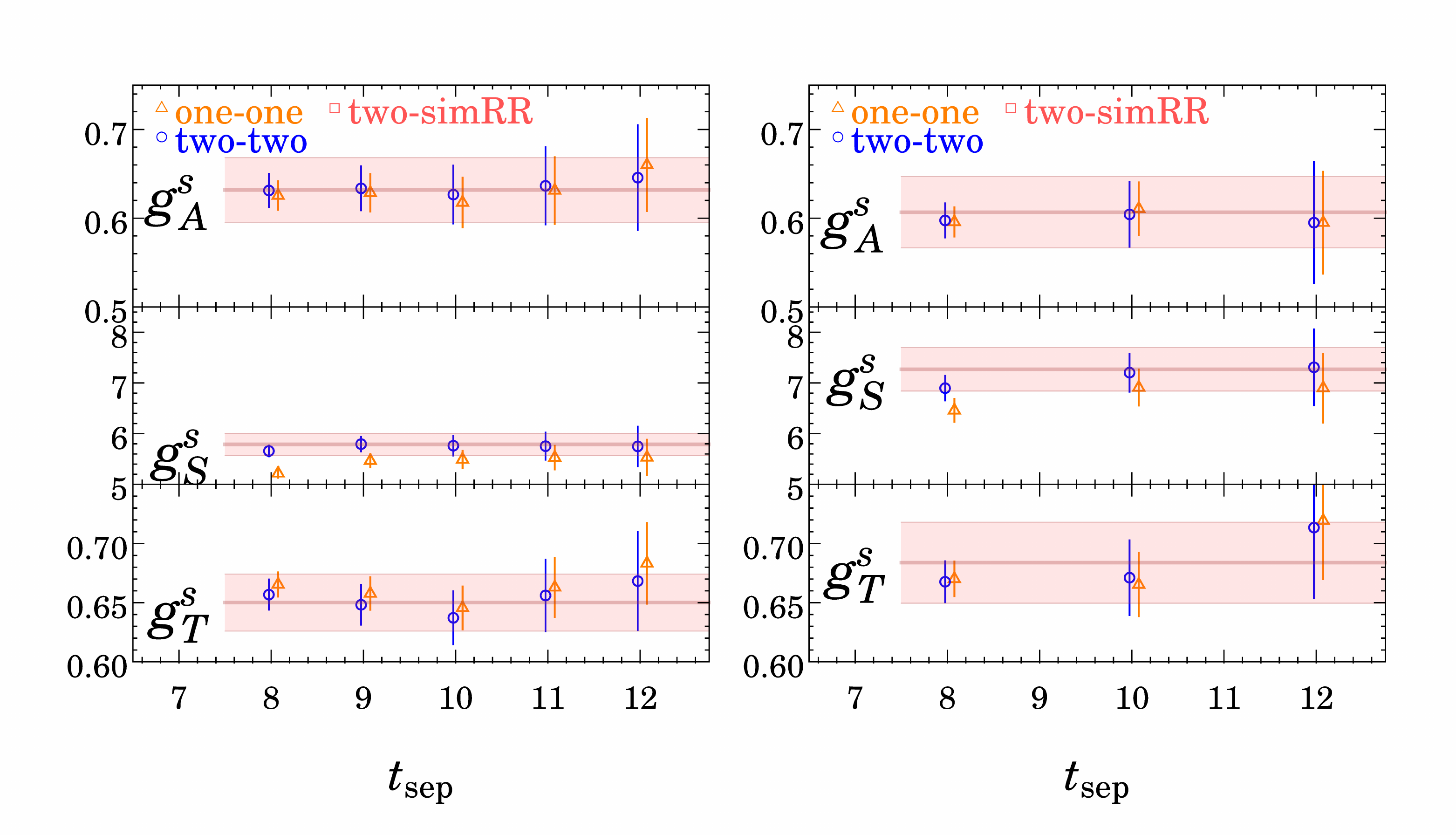}
\caption{The bare isoscalar charges: $g_{A,S,T}^{s}$ from top to bottom for $M_\pi\approx$ 310 (left column) and 220 (right column) MeV as functions of source-sink separation $t_\text{sep}$ (in lattice units).
Estimates are shown for the four different fit methods described in the text. The band shows the results of the two-simRR fit to data for all $t_\text{sep}$.
}
\label{fig:isoscalar}
\end{figure*}

In Fig.~\ref{fig:isoscalar}, we show estimates of the unrenormalized
connected parts of the isoscalar axial, scalar and tensor charges,
$g_{A,S,T}^s$, extracted using the same analysis methods described
in Sec.~\ref{sec:excited}. Again, we find consistency between the
methods for our tuned smearing-parameter choices. The one
exception is the isoscalar scalar charge, for which estimates based on
one-one or ratio methods show an increase with the source-sink separation
for $t < 10$. Overall, the two-simRR method gives our best estimates, and
these agree with those from the other methods for $t_\text{sep} \ge 10$.

The renormalization constants for the isoscalar charges also receive
contributions from disconnected diagrams. In the approximation that
disconnected diagrams are neglected, the renormalization constants for
the isoscalar and isovector charges are the same. We, therefore, use
results in Table~\ref{tab:resultsZ} to renormalize the connected
isoscalar charges at 2~GeV in $\overline{\text{MS}}$ scheme. These
estimates of the renormalized charges are given in
Table~\ref{tab:isoscalar}.  We also include estimates for the value at
the physical pion mass using a linear extrapolation in $M_\pi^2$.

\begin{table*}
\begin{tabular}{|c|ccc|ccc|}
\hline
              &  $g_A^{s,\text{bare}}$ &  $g_S^{s,\text{bare}}$ &  $g_T^{s,\text{bare}}$  &  $g_A^s$     &  $g_S^s$    &  $g_T^s$      \\\hline
310-MeV       & $0.632(36)$          & $5.79(22) $          & $0.650(24)$           &  $0.610(37)$   &  $5.16(24)$   & $0.613(26)$     \\\hline
220-MeV       & $0.607(40)$          & $7.27(43) $          & $0.684(34)$           &  $0.579(39)$   &  $6.40(41)$   & $0.636(35)$     \\\hline\hline
Extrapolation &                      &                      &                       &  $0.559(67)$   &  $7.15(65)$ & $0.651(58)$     \\\hline
\end{tabular}
\caption{The unrenormalized (left half) and renormalized (right half) estimates for the connected parts of the isoscalar charges. 
The renormalization constants used to convert these to the $\overline{\text{MS}}$ scheme 
at 2~GeV are the same as given in Table~\ref{tab:resultsZ}.  
The error quoted on the renormalized charges are obtained by combining the statistical
and systematic errors in the ratios of $Z$'s and $g$'s in quadrature under
the assumption that they are independent. 
The extrapolation of the renormalized charges to 
$140$~MeV is carried out using a fit linear in $M_\pi^2$. 
}
\label{tab:isoscalar}
\end{table*}

\section{Isovector Electromagnetic Form Factors of the Nucleon}
\label{sec:formfactors}

The Dirac and Pauli form factors ($F_{1,2}$) are extracted from the
matrix elements of the isovector vector current in the nucleon state
$N$ through the relation
\begin{multline}
\label{eq:Vector-roper}
\left\langle N(\vec{p}_f) | V_\mu (\vec{q}) | N(\vec{p}_i)\right\rangle  = \\
{\overline u}_N(\vec{p}_f)\left( F_1(Q^2) \gamma_\mu
+\sigma_{\mu \nu} q_\nu
\frac{F_2(Q^2)}{2 M_N}\right)u_N(\vec{p}_i),
\end{multline}
where the momentum transfer $q=p_f-p_i$.
Another common set of definitions of these form factors, widely used in
experiments, are the Sachs (electric and magnetic) form factors;
which can be related to the Dirac and Pauli form factors through
\begin{align}\label{eq:sachs}
G_E(Q^2) &= F_1(Q^2) - \frac{Q^2}{4M_N^2}F_2(Q^2) \\
G_M(Q^2) &= F_1(Q^2) + F_2(Q^2).
\end{align}

A compilation of the lattice parameters used by various collaborations
performing simulations of the electromagnetic form factors is given
in Table~\ref{tab:emffothers}.

\begin{table*}
\footnotesize
\setlength{\tabcolsep}{1pt}
  \begin{tabular}{|lccccccc|}
\hline
Collaboration                   & Action             & $N_f$ & $M_\pi$ (MeV) & $L$ (fm) & $(M_\pi L)_\text{min}$ & $a$ (fm)    &   Observables Calculated  \\
\hline\hline
RBC\cite{Lin:2008uz}            & DWF                & 2     & 490--695 & 1.9           & $4.75$  & 0.117                     & $F_{1,2}^v$, $\kappa^v$, $(r_{1,2}^v)^2$ \\
ETMC\cite{Alexandrou:2011db}    & twisted Wilson     & 2     & 260--470 & \{2.1, 2.8\}  & $3.3$   & $\{0.056,0.070,0.089\}$   & 
$F_{1,2}^v$, $\kappa^v$, $(r_{1,2}^v)^2$\\
CLS-Mainz\cite{Capitani:2012ef}      & clover             & 2     & 277--649 & 2.0--3.0      & $4.0$   & $\{0.05,0.06,0.08\}$      & $G_{E,M}^v$, $(r_1^v)^2$ \\
\hline
Lin+Orginos\cite{Lin:2008mr}    & DWF on staggered   & 2+1   & 354-754  & 2.5           & $3.68$  & 0.1224                    & $(r_{E,M})^2$, $\mu^{p,n}$, $G_{E,M}^{p,n}$, \\
LHPC\cite{Bratt:2010jn}         & DWF on staggered   & 2+1   & 290--870 & $\{2.5,2.7\}$ & $3.68$  & 0.1224                    & $F_{1,2}^v$,$(r_{1,2}^v)^2$, $\kappa^v$, $G_{E,M}^v$  \\
HSC\cite{Lin:2010fv}            & anisotropic clover & 2+1   & 450--840 & 2.0           & $4.57$  & 0.125 ($a_t=0.036$)       & $F_{1,2}^v$,$(r_{1,2}^v)^2$, $\kappa^v$, $G_{E,M}^{p,n}$  \\
RBC/UKQCD\cite{Yamazaki:2009zq} & DWF                & 2+1   & 330--670 & $\{1.8,2.7\}$ & $3.8$   & 0.114                     &  $F_{1,2}^v$,$(r_{1,2}^v)^2$, $\kappa^v$  \\
LHPC\cite{Green:2012ud}         & clover             & 2+1   & 149--356 & $\{2.8,5.6\}$ & 3.57                                & \{0.116, 0.09\} &$\kappa^v$,  $(r_{1,2}^v)^2$ \\
\hline
ETMC\cite{Alexandrou:2013joa}   & twisted Wilson     & 2+1+1 & 354--465  & $\{2.5,2.9\}$ & 3.35   & $\{0.066,0.086\}$       & $G_{E,M}^v$, $F_{1,2}^v$, $\kappa^v$, $(r_{1,2}^v)^2$\\
PNDME (this work)               & clover on HISQ     & 2+1+1 & 220--310 & \{2.88,3.84\} & 4.28    & 0.12                      & $F_{1,2}^v$,$(r_{1,2}^v)^2$, $\kappa^v$ \\
\hline
\end{tabular}
\caption{A summary of the lattice parameters used by collaborations
carrying out calculations of the nucleon electromagnetic form factors.}
\label{tab:emffothers}
\end{table*}

\begin{figure*}
\includegraphics[width=0.35\textwidth]{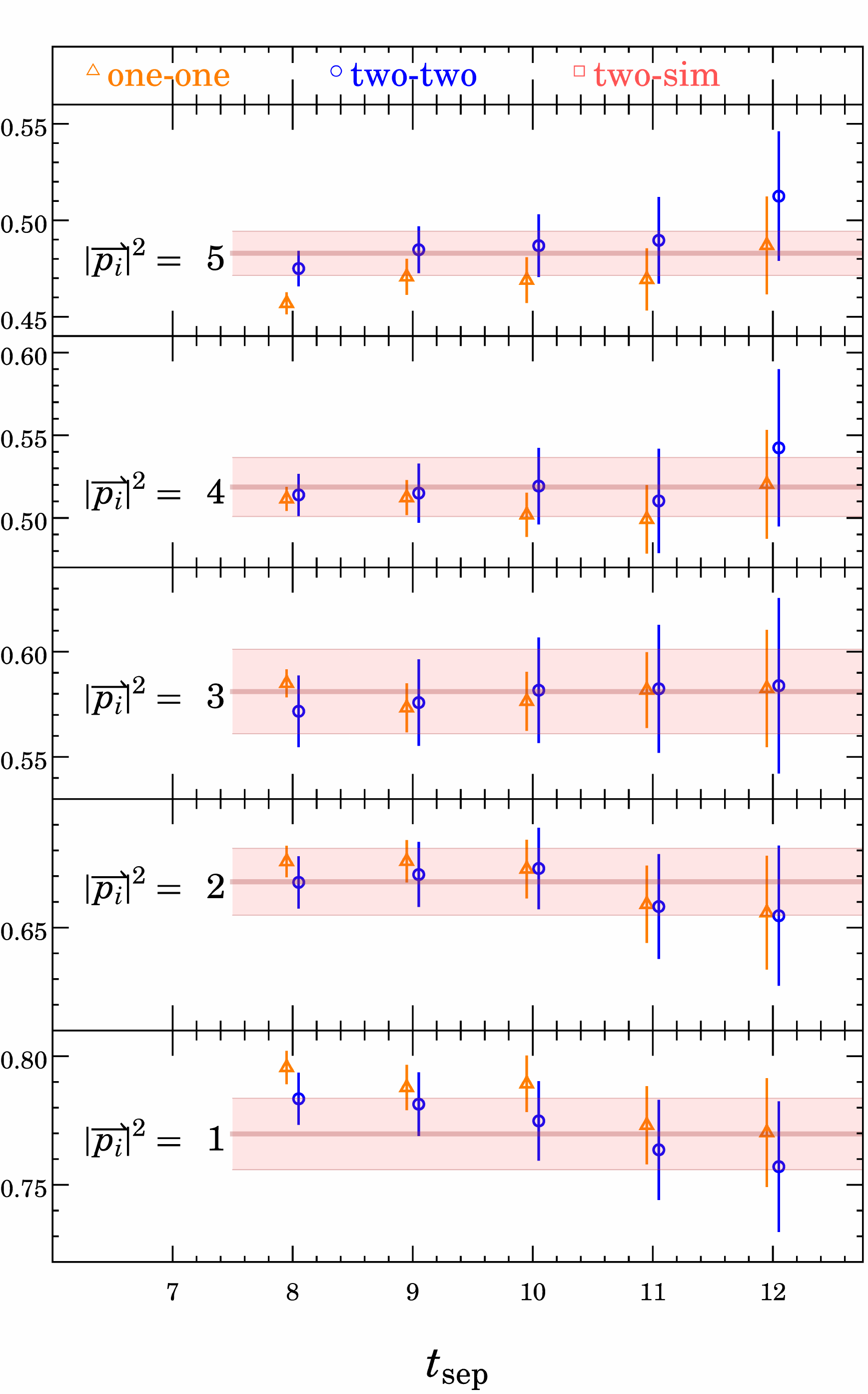} \hspace{1cm}
\includegraphics[width=0.35\textwidth]{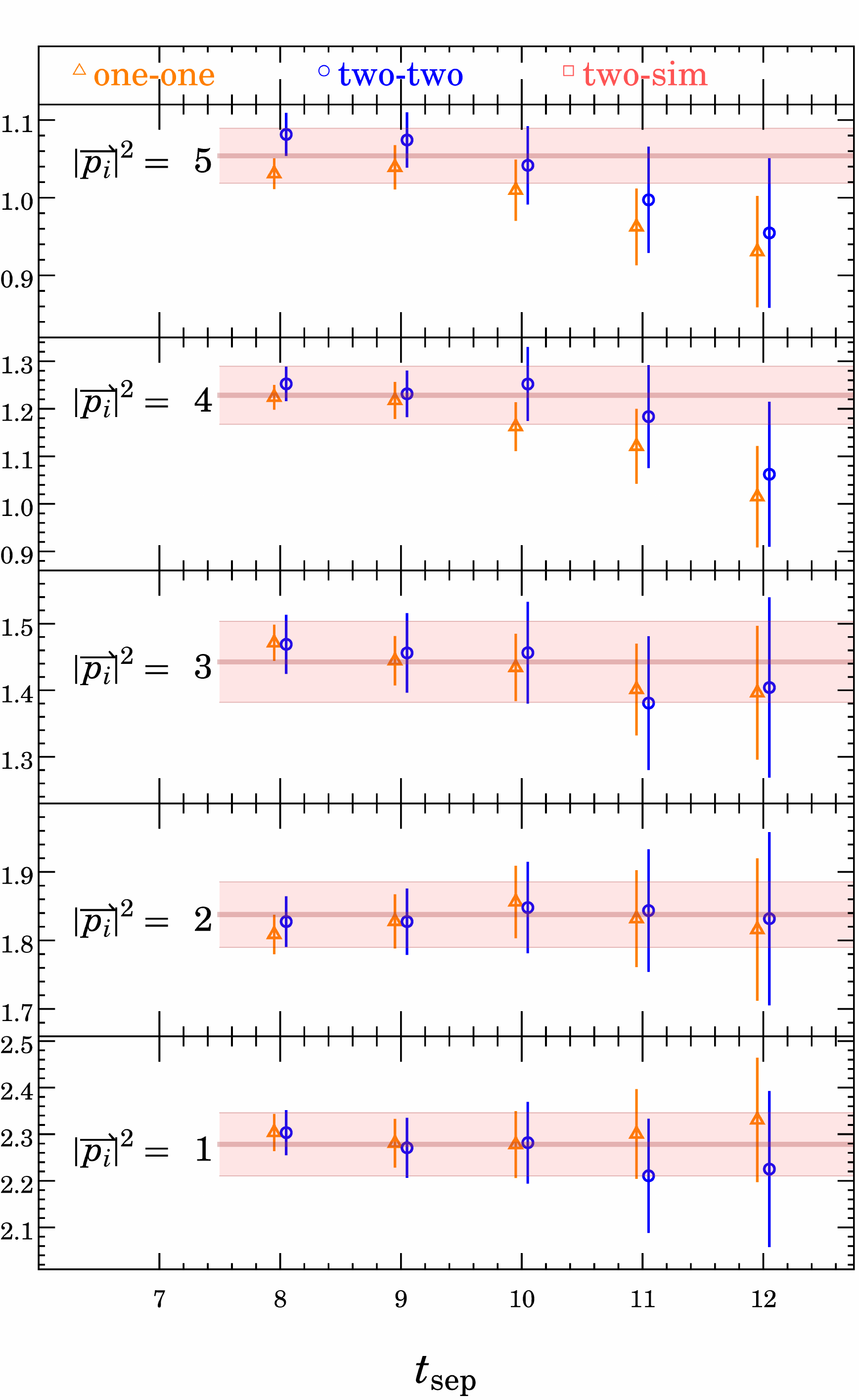}
\\
\includegraphics[width=0.35\textwidth]{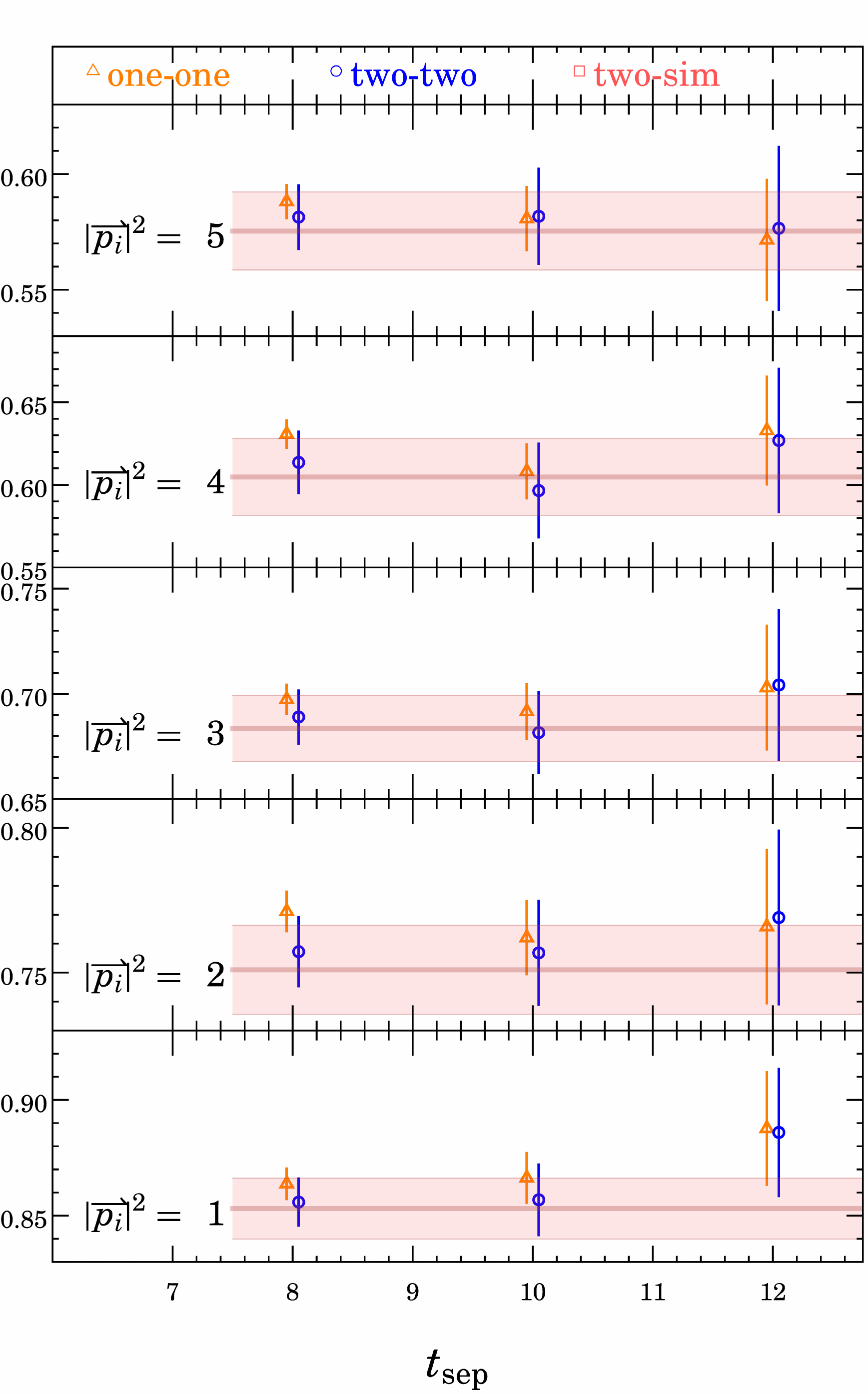} \hspace{1cm}
\includegraphics[width=0.35\textwidth]{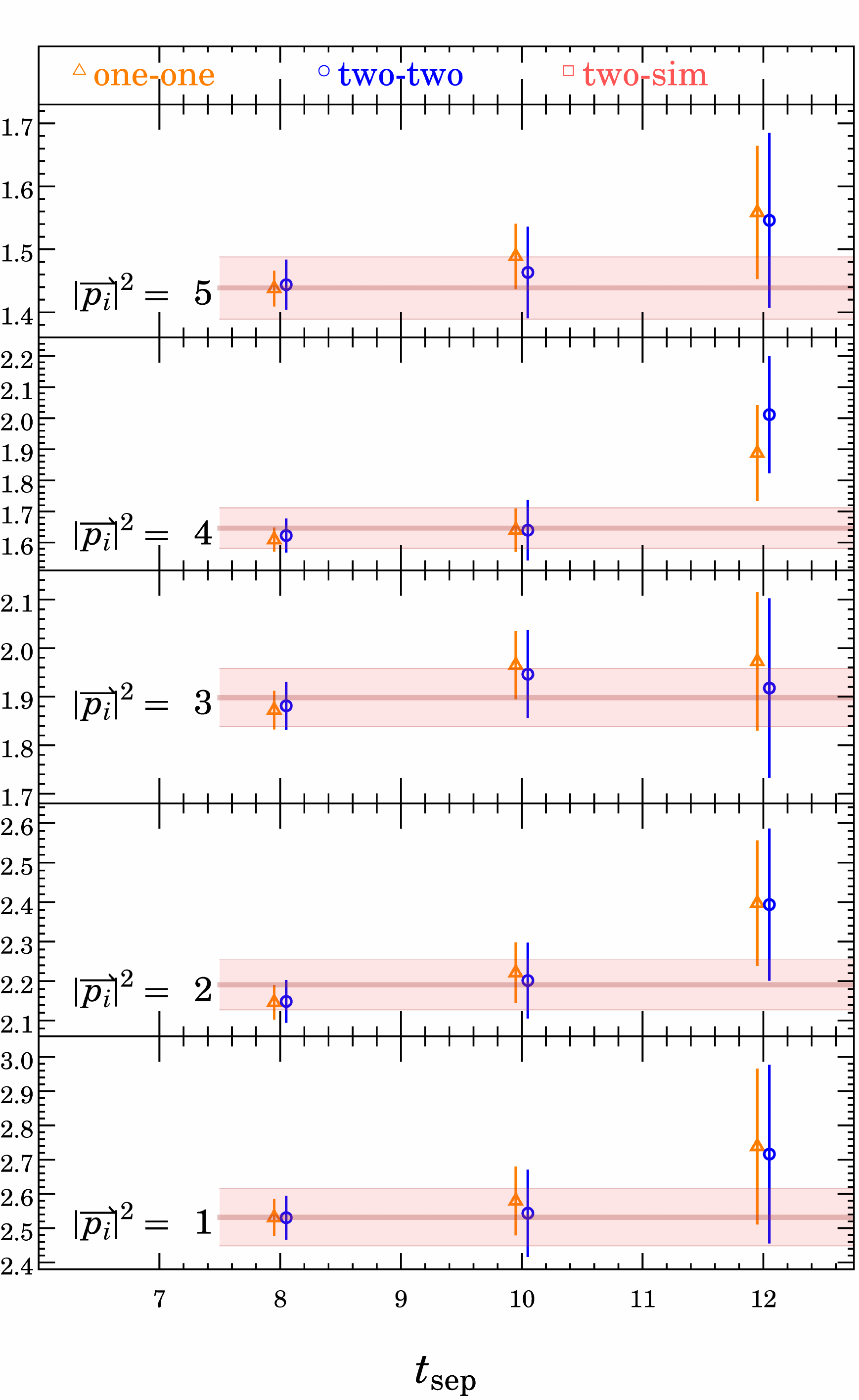}
\caption{(top) The data for the isovector Dirac (left) and Pauli (right) form
  factors $F_{1,2}^v$ (normalized by $F_1^v(Q^2=0)$) for the 310-MeV ensemble for
  all momenta. Individual data points in the figures are extracted using the
  one-one and two-two methods, while the bands are values using a two-sim fit to all
  values of $t_\text{sep}$ calculated. (bottom) The same form factors obtained on 220 MeV ensembles.}
\label{fig:formfactors}
\end{figure*}

The vector-current matrix elements, $\left\langle
N\middle|V_\mu\middle|N\right\rangle$ in the ground state (with $n=n^\prime=0$)
at different momenta are obtained by using the same projection
matrices $T$ as in Eq.~\ref{eq:general-three-point}. This overdetermined system
of linear equations allows solution for the Dirac and Pauli form
factors $F_{1,2}$ with various Lorentz indices $\mu$ and momenta $\vec p_i$ for a particular
$Q^2$. To minimize the excited-state contribution to the ground-state
matrix element, we again employ a number of fits, using the notation
established in Sec.~\ref{sec:excited}.
The upper two plots of Fig.~\ref{fig:formfactors} show 310-MeV Dirac and Pauli form
factor at each $Q^2$ as a function of $t_\text{sep}$. Once again, we observe 
that fitted values from two-two method are consistent with those from two-sim fit, while
  one-one method is less consistent.
The lower two plots of Fig.~\ref{fig:formfactors} show examples from 
220-MeV ensemble for all values of $t_\text{sep}$ investigated.
We adopt as our preferred value the two-sim fit, which takes into account excited-state systematics.
Relative to our adopted values, we find the central values for the one-one and two-two method
shift by no more than $2\sigma$ with our statistics with the smallest separation.
We find that the differences with and without the 
$\langle 1 | O_\Gamma |1 \rangle$ term in Eq.~\ref{eq:general-three-point} 
are consistent with zero within errors, and including the RR term
increases the errors. Therefore, the final
results presented use the two-sim fit neglecting this term. 

We normalize the Dirac and Pauli form factors $F_{1,2}$ by the value
$F_1^v(Q^2=0)$ determined directly in the calculation; thus, the renormalization
factor $Z_V$ cancels, and the $O(a)$-systematics are reduced. In the case of
$F_1$, we explore two functional forms to characterize the $Q^2$ behavior: a
conventional dipole $(1+Q^2/M^2)^{-2}$ with one free parameter and a more
general quadratic in $Q^2$, $(1+b_1 Q^2 + b_2 Q^4)^{-1}$ with two free
parameters. We find that dipole form does not work for all of our form factor
data at any source-sink separation; we have to cut the data as low as $Q^2 \leq
0.4\mbox{ GeV}^2$ to make it work. Since there is no fundamental physics reason
for using this form, we take the central value from the general quadratic form,
which gives a much better fit. Unfortunately, including more free parameters in
the fit results in the final extrapolated value of the charge radii having
larger uncertainty.
In the case of $F_2$, we also investigated multiple fit ans\"atze: (i)~dipole
$F_2^v(0)(1+Q^2/M^2)^{-2}$; (ii) tripole $F_2^v(0)(1+Q^2/M^2)^{-3}$ and (iii)~a
general form $F_2^v(0)(1+c_1 Q^2 + c_3 Q^6)^{-1}$, with the anomalous magnetic moment
$\kappa^v \equiv F_2^v(0)$. We find that all three ans\"atze capture the data
reasonably, since the errorbars are larger in Pauli form factor. However, we
choose to use the general ansatz for the final fit.

\begin{figure*}
\includegraphics[width=0.45\textwidth]{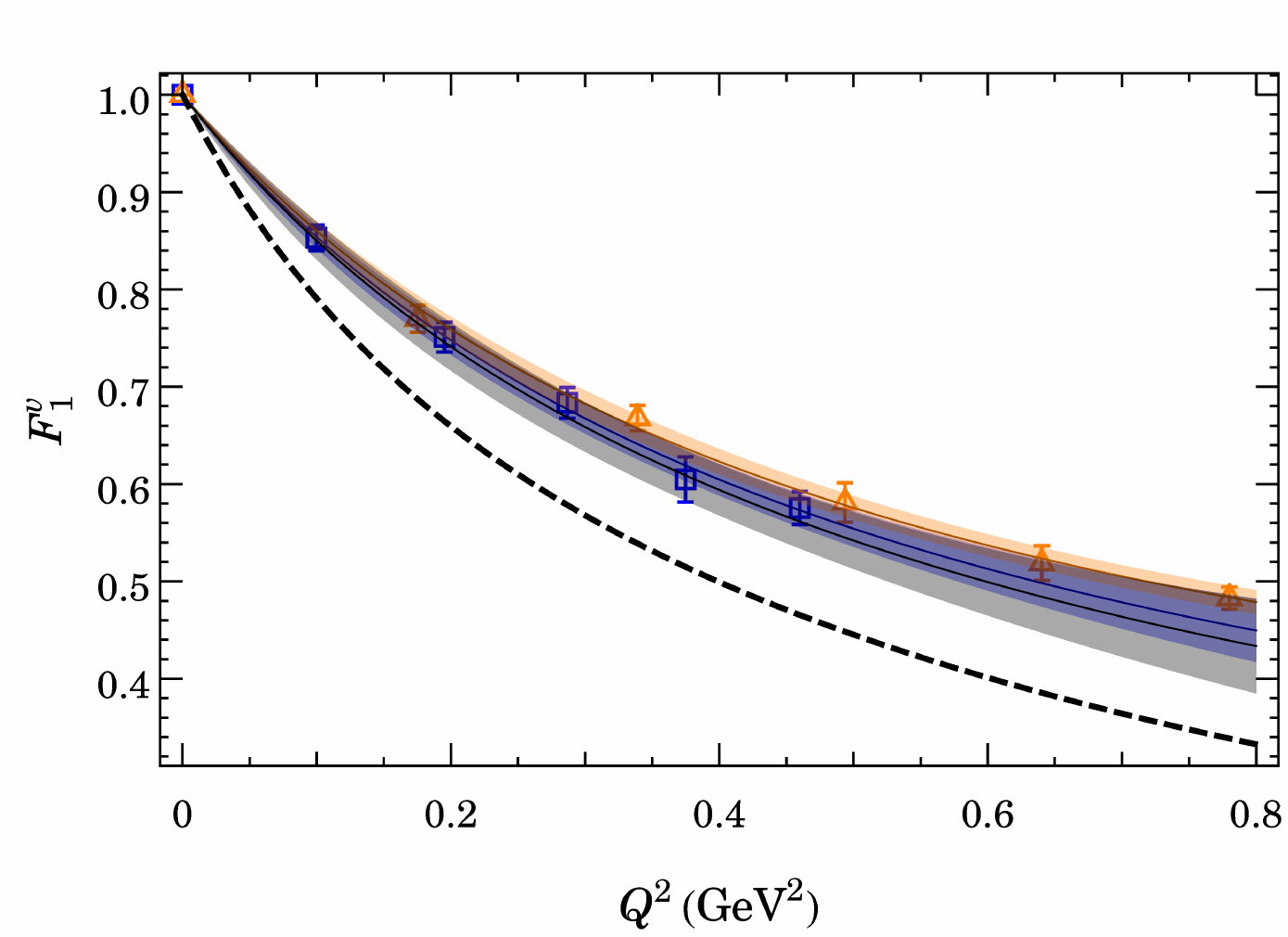}
\includegraphics[width=0.45\textwidth]{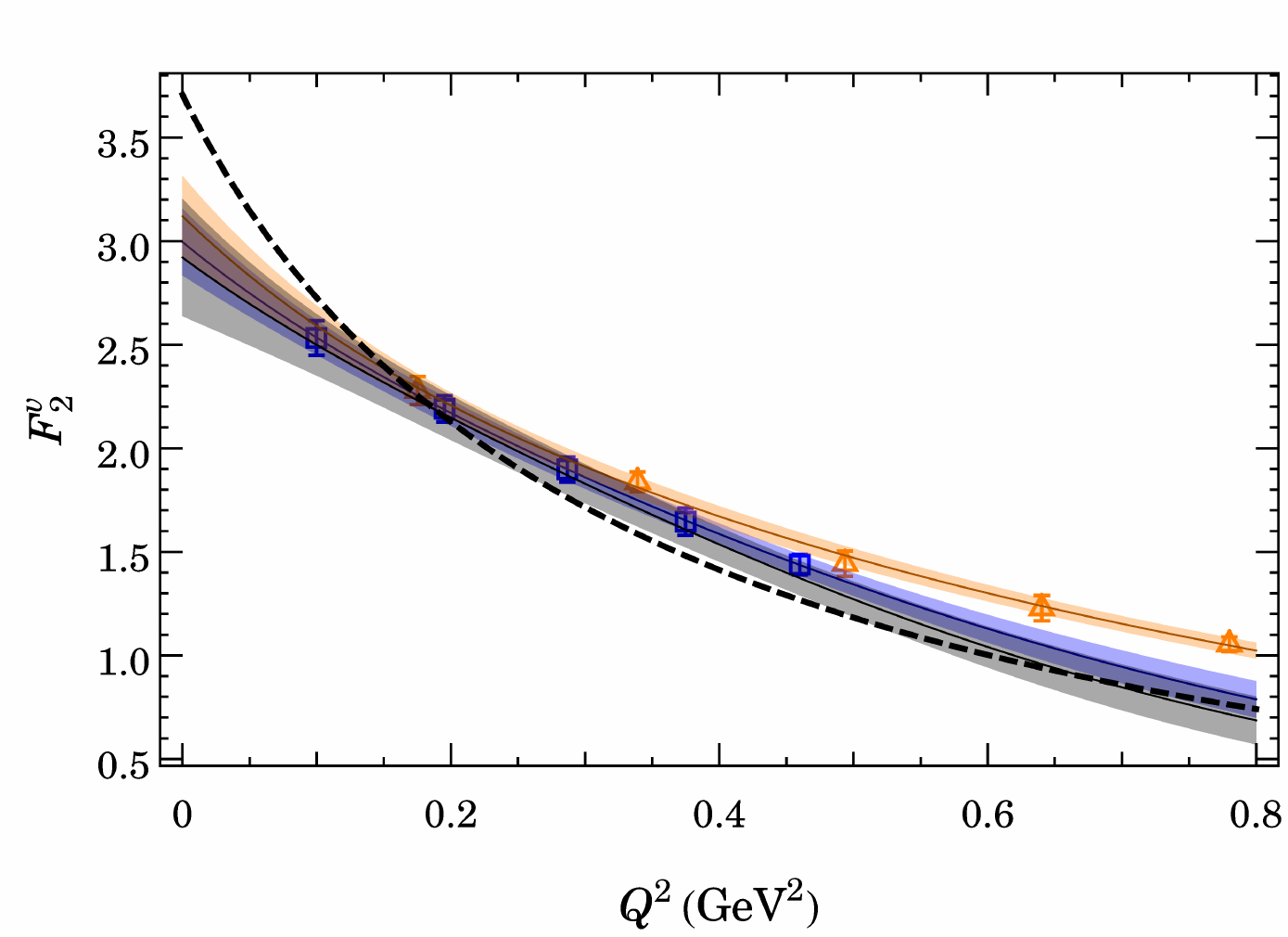}
\includegraphics[width=0.45\textwidth]{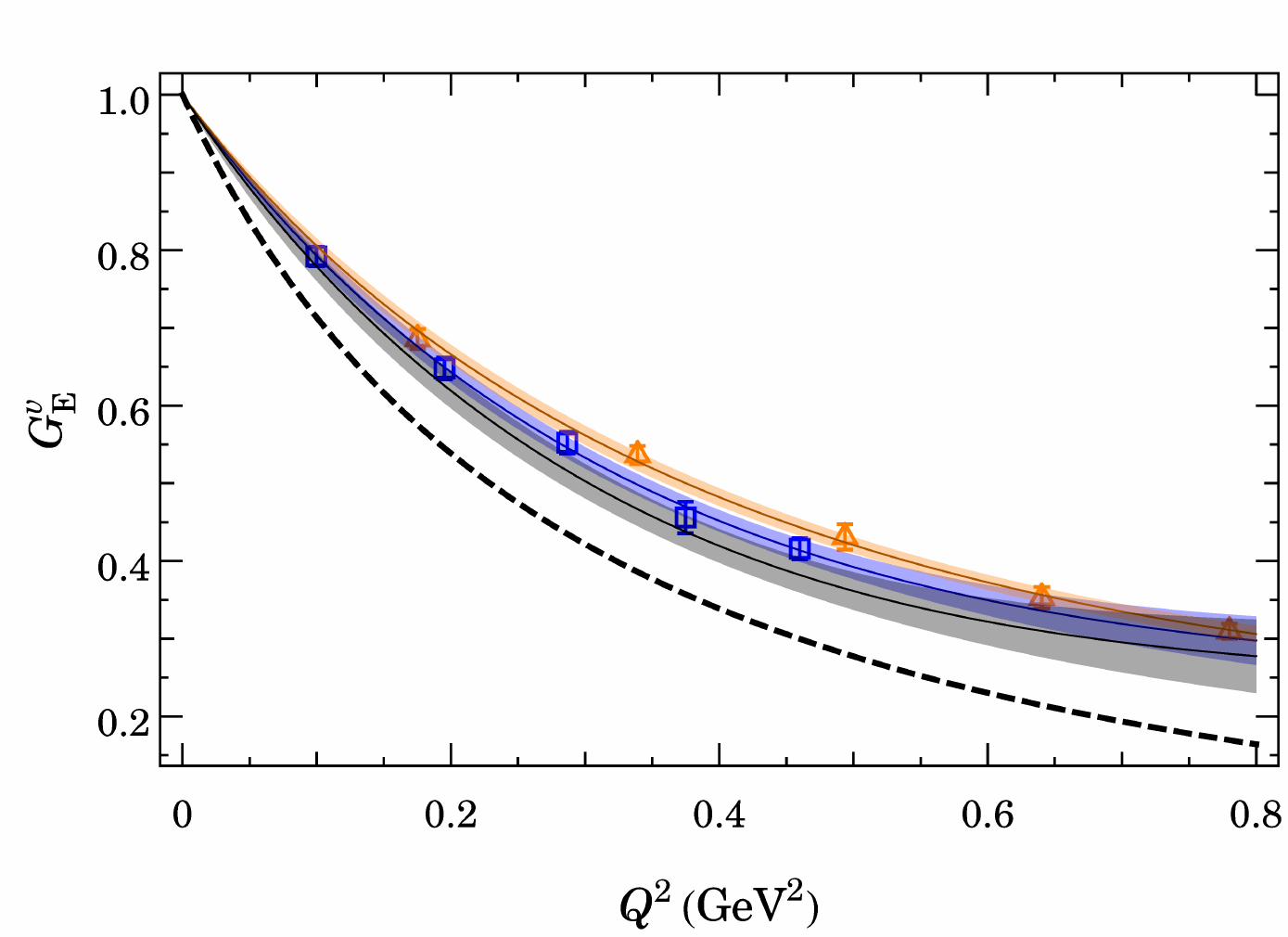}
\includegraphics[width=0.45\textwidth]{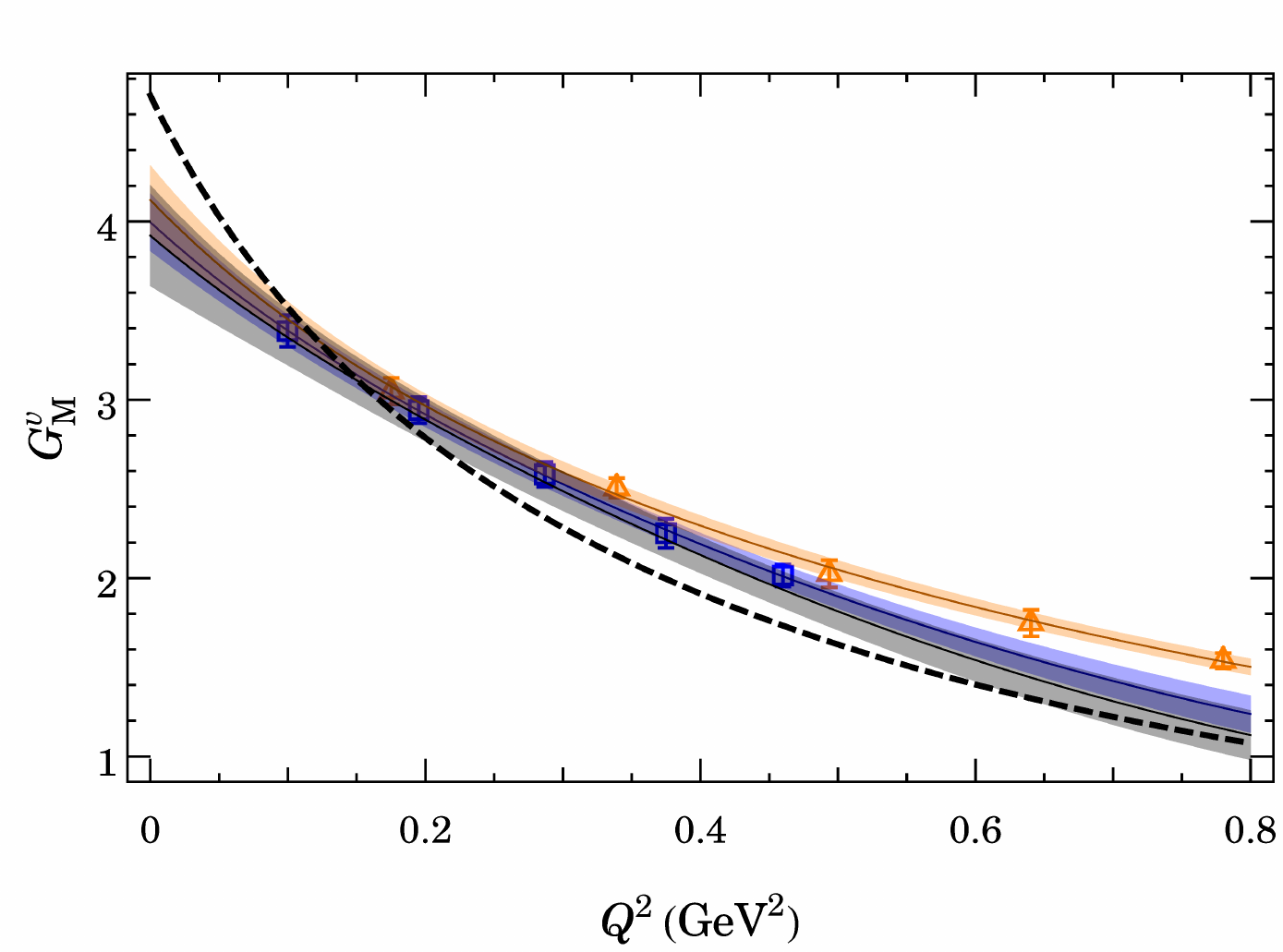}
\caption{(Top) The data for the isovector Dirac and Pauli form factors
  $F_{1,2}^v$ (normalized by $F_1^v(Q^2=0)$) for the $M_\pi=310$ MeV (triangles)
  and $M_\pi=220$ MeV (squares) ensembles from two-sim fit. The bands
  are the general form fit through all $Q^2$. The lower darker bands are the
  extrapolations to physical pion mass, and the dashed curves are the experimental
  parametrization~\cite{Arrington:2007ux,Alberico:2008sz}. (bottom) The
  same data and fits shown in terms of the Sachs electric and magnetic form
  factors $G_{E,M}^v$.}
\label{fig:F12fits}
\end{figure*}

Figure~\ref{fig:F12fits} shows both 310 and 220-MeV Dirac and Pauli form-factor
results with the general ansatz, and the extrapolation to the physical pion
mass point. We found small pion-mass dependence on these ensembles. For Dirac
form factors, our form factors are larger than the experimentally reconstructed values;
consequently, the smaller slope around the $Q^2=0$ point gives smaller charge
radius as defined below. This feature has been observed in the past with pion
mass larger than 300~MeV. The disagreement in the Pauli form factors is less
severe, but there is also misalignment in the small-$Q^2$ region.

The size of the nucleon characterized by the effective Dirac and Pauli
radii can be determined from the electromagnetic form
factors. These are determined from the slope of the
corresponding form factor in the zero-$Q^2$ limit.
\begin{equation}\label{eq:GEradii}
\langle r_{1,2}^2\rangle = -6\frac{d}{dQ^2}\left.\left(\frac{F_{1,2}^v(Q^2)}{F_{1,2}^v(0)}\right)\right|_{Q^2=0}.
\end{equation}
Since the value of the smallest momenta allowed in typical lattice
simulations is large, to extract the radii it is important
to develop ans\"atze that capture the low-$Q^2$ behavior well. We use
the general ansatz defined above to extract the radii, since they give the
best fit. Attempts are being made to obtain data at smaller momenta
to improve the determination~\cite{Hagler:2009ni}.

\begin{figure*}
\includegraphics[width=0.45\textwidth]{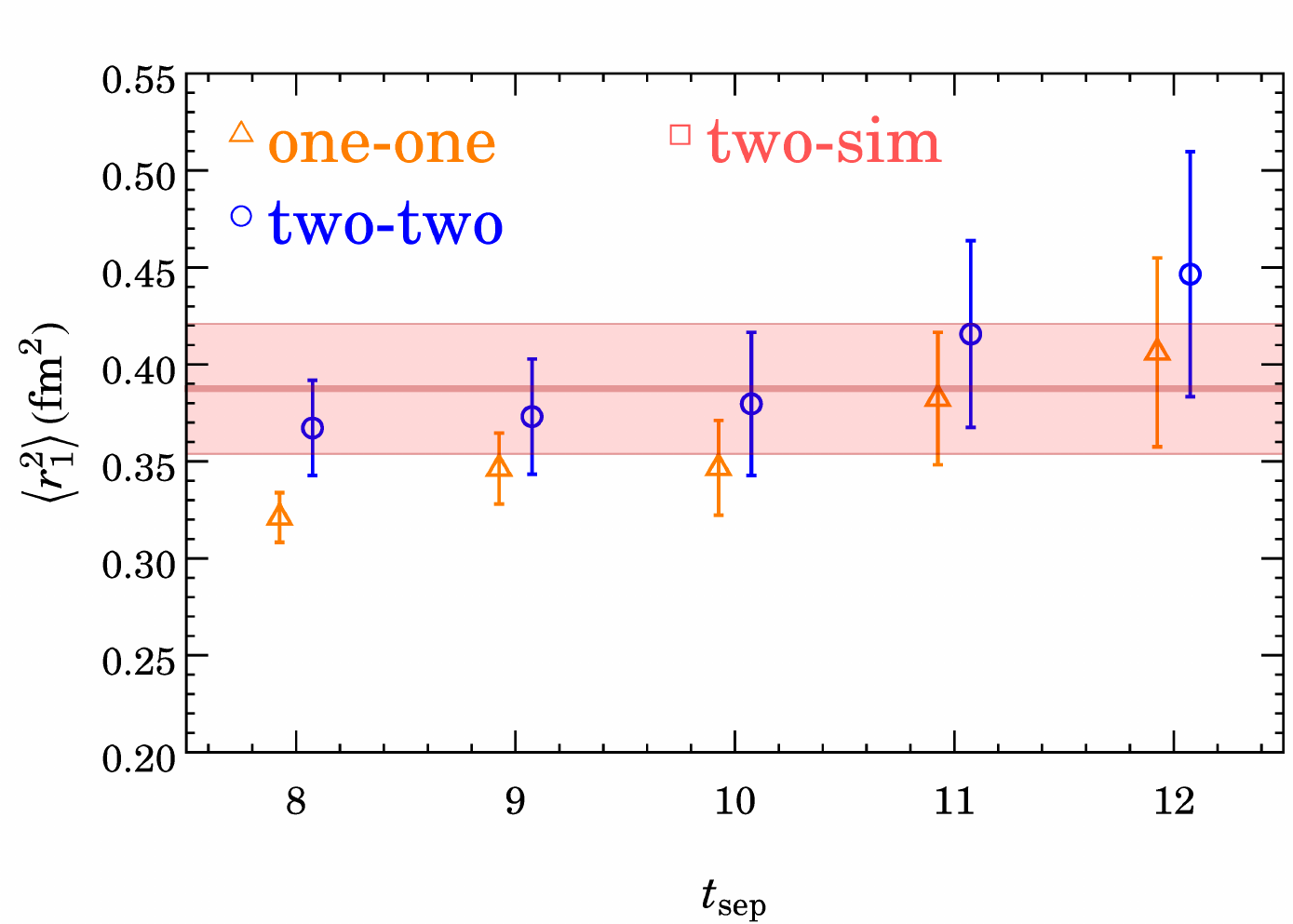}
\includegraphics[width=0.45\textwidth]{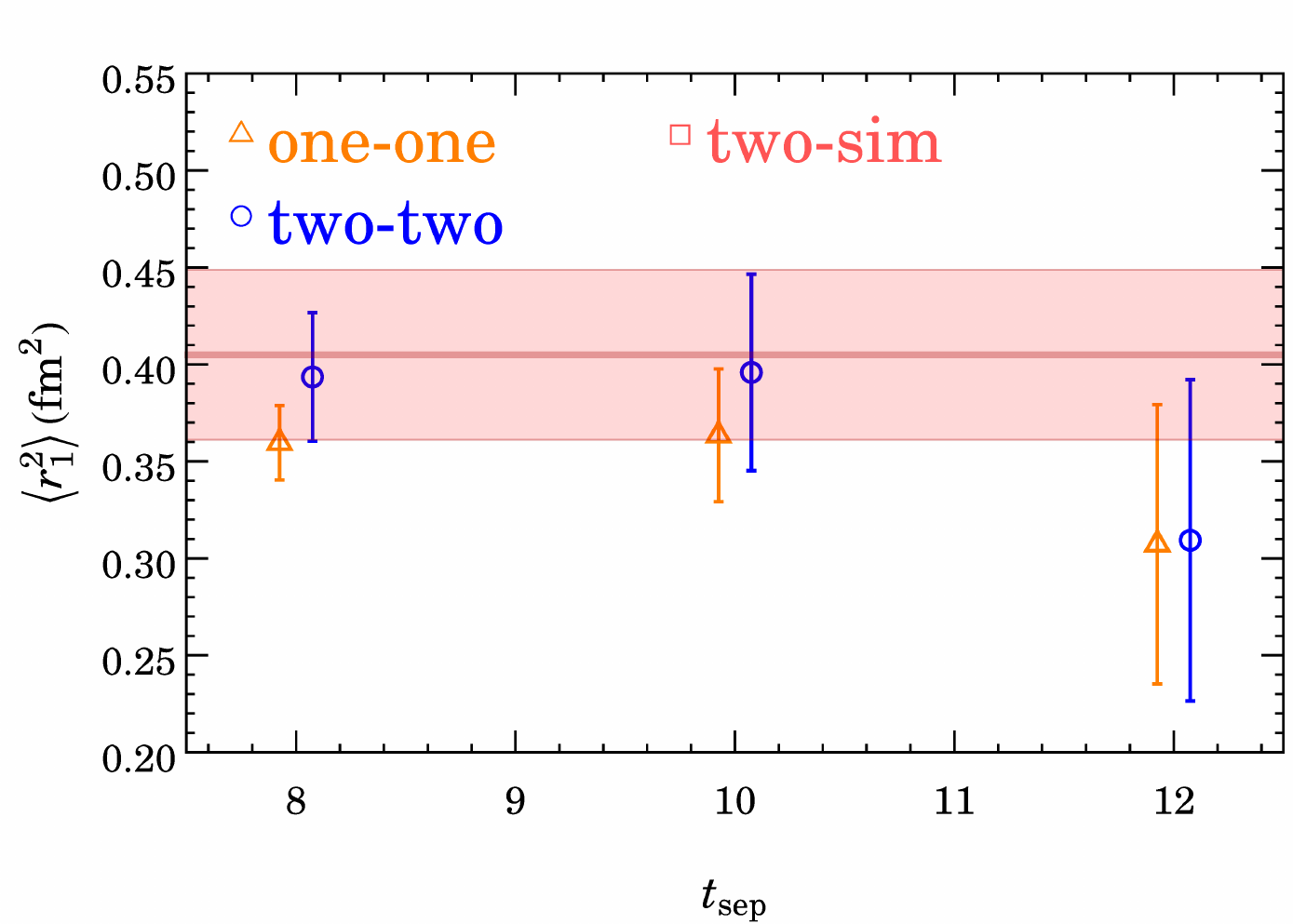}
\includegraphics[width=0.45\textwidth]{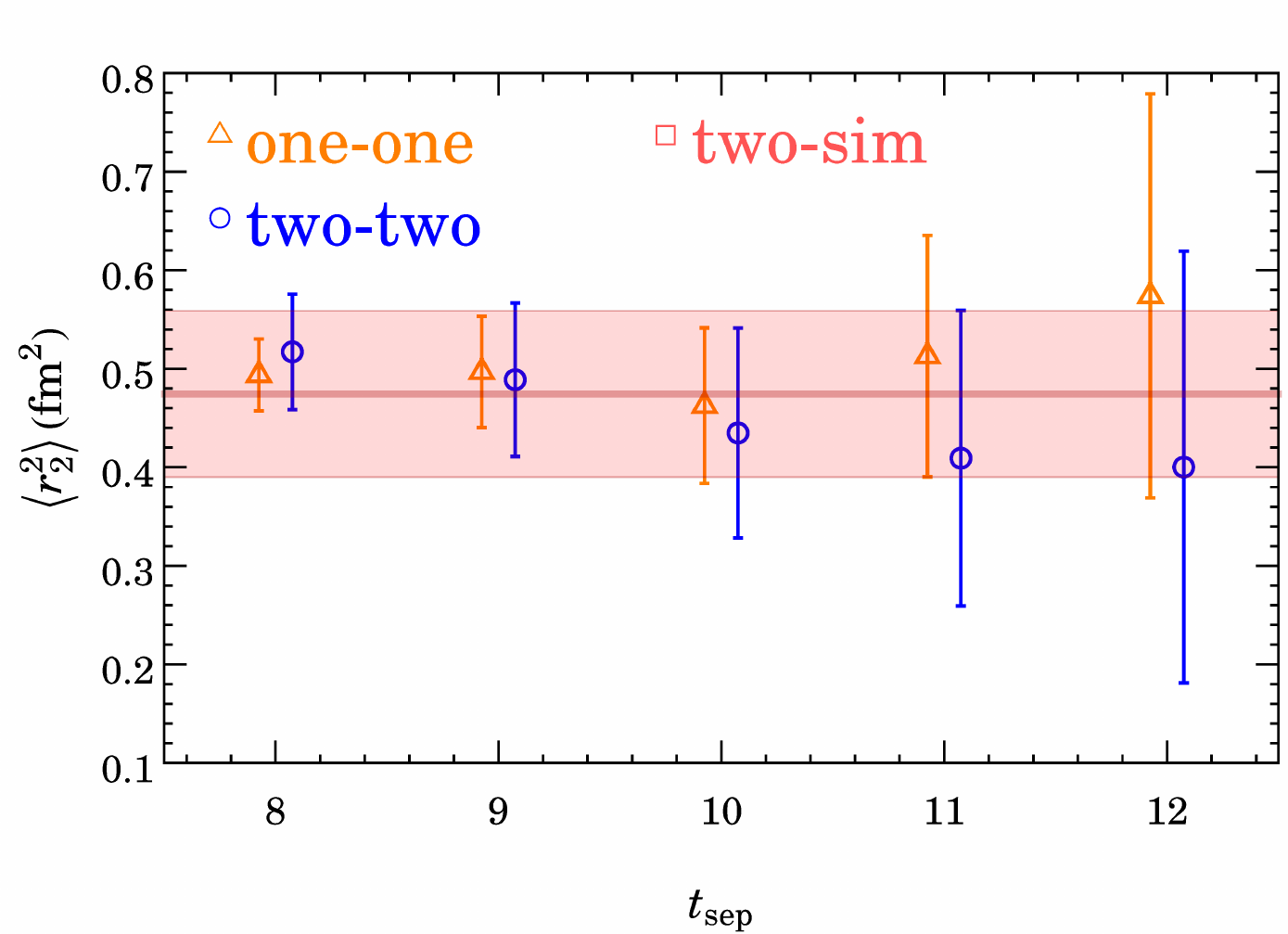}
\includegraphics[width=0.45\textwidth]{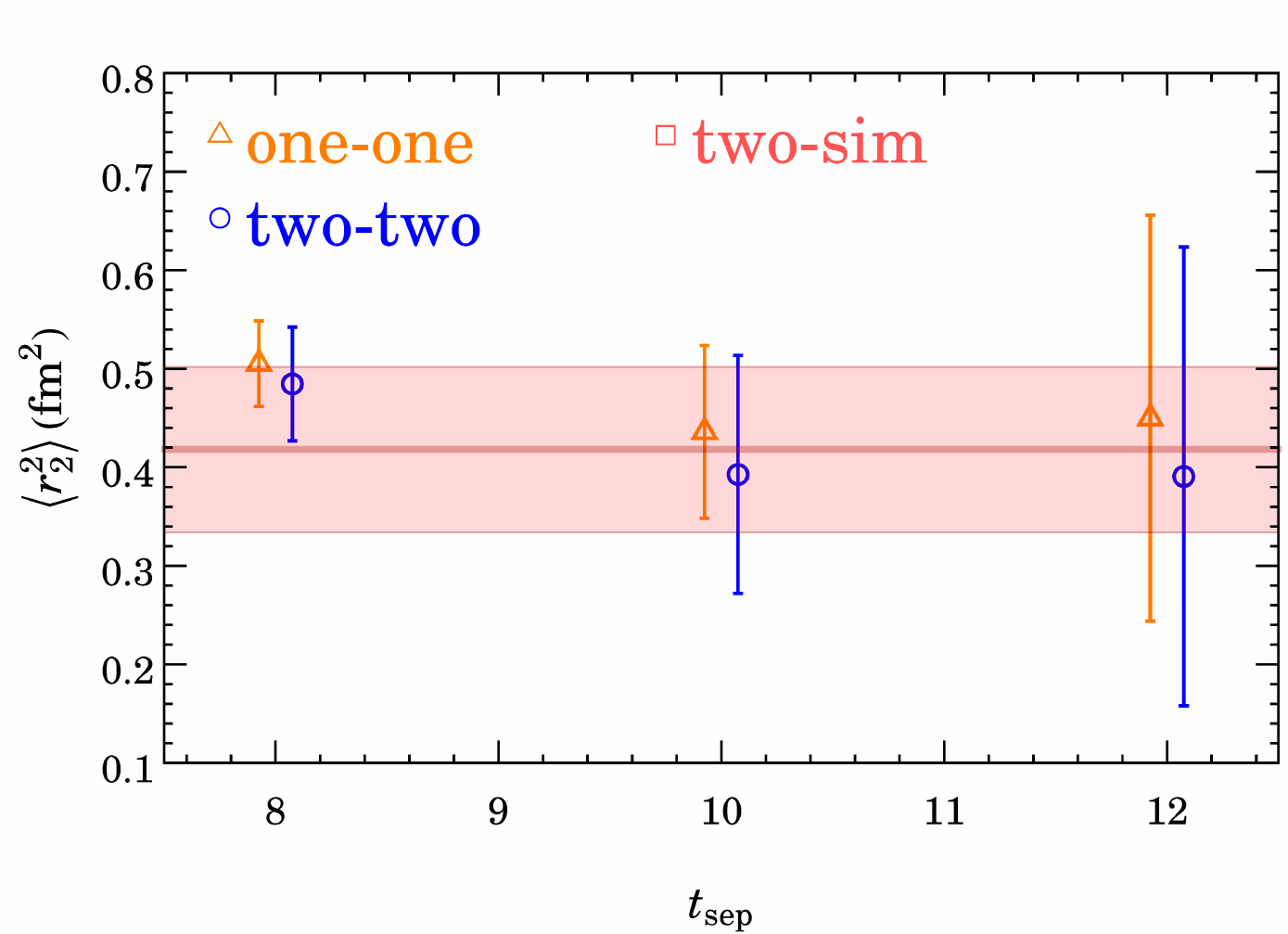}
\includegraphics[width=0.45\textwidth]{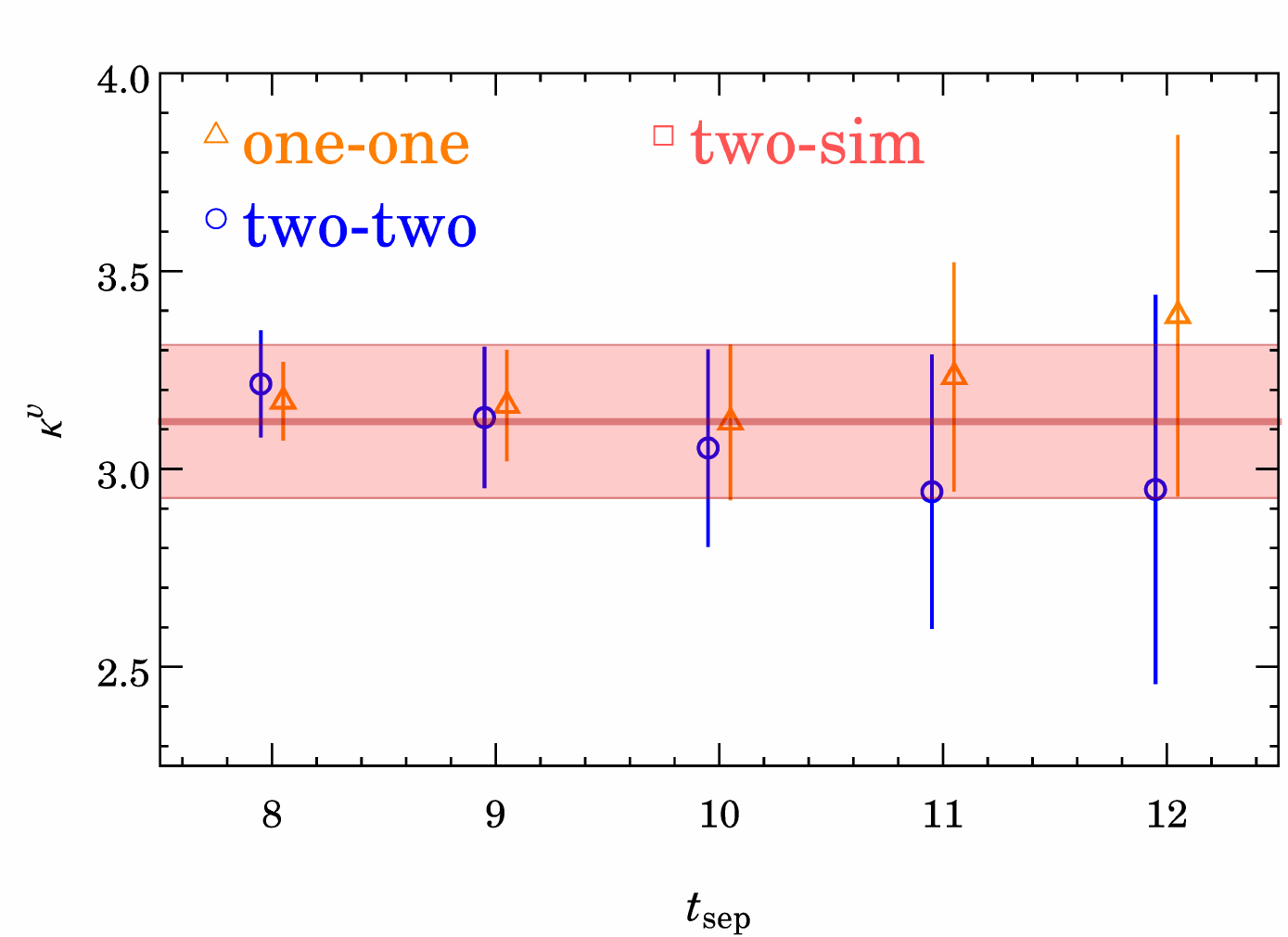}
\includegraphics[width=0.45\textwidth]{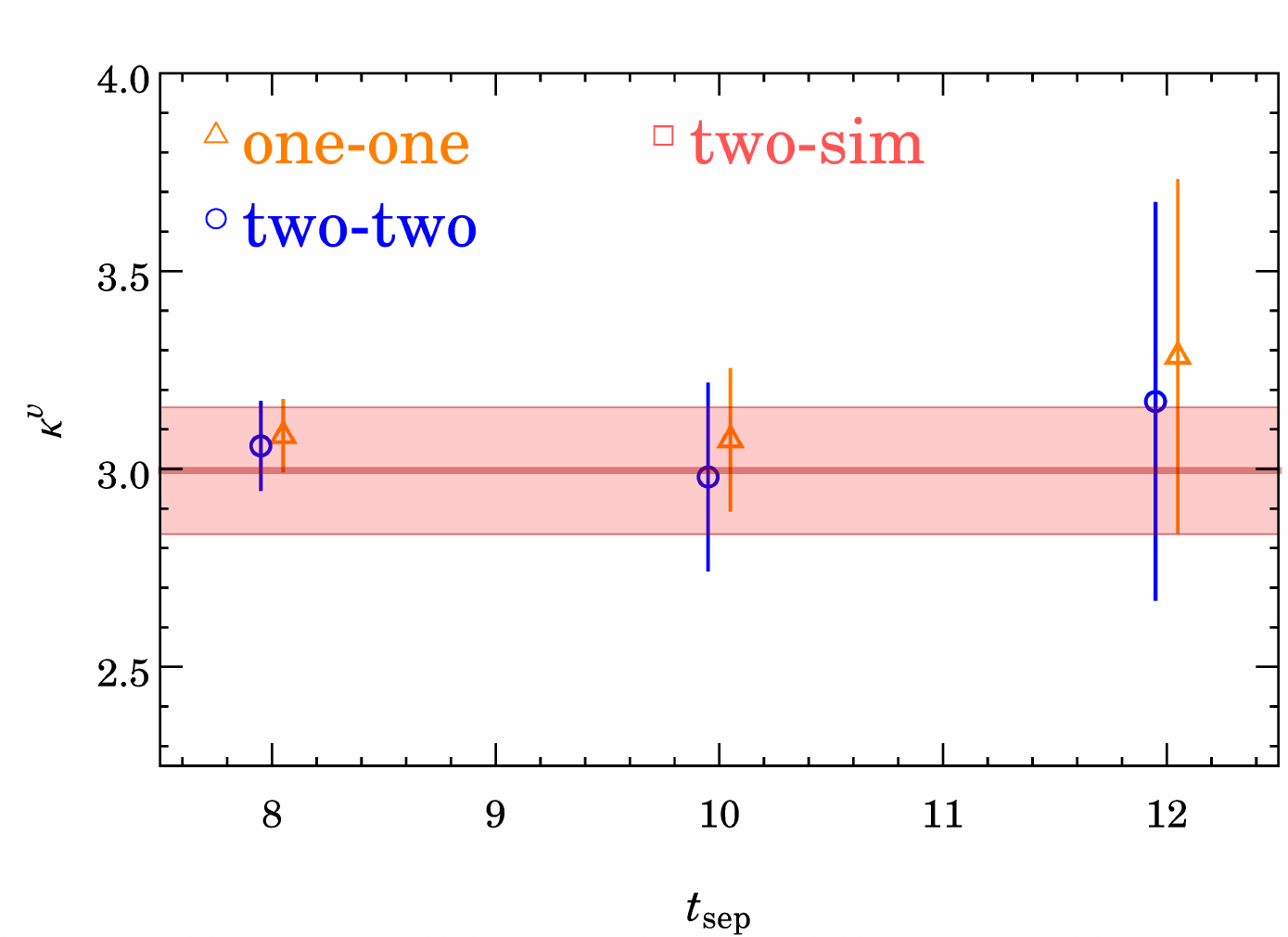}
\caption{Data for the isovector Dirac (top) and Pauli (middle) radii, and
  anomalous magnetic moments (bottom) from $M_\pi=310$ (left) and 220 (right)
  MeV ensembles as functions of source-sink separation and three analysis
  methods: one-one (triangles), two-two (circles) and two-sim (band). }
\label{fig:r12}
\end{figure*}

Figure~\ref{fig:r12} shows our results for the Dirac and Pauli radii
for three fit methods and the two ensembles. In the case of the Dirac
radii, the one-one method on the 310-MeV ensemble data becomes reliable
only at larger separation; the central values increase (within
errorbar) with separation.
We find that using the two-two method, which
includes the leading Roper-nucleon contribution, gives consistent
radii for all source-sink separations. The 220-MeV data do not show
any significant trend with $t_\text{sep}$. In the
case of Pauli radii, we do not observe significant dependence on the
separation nor on the analysis method. Similar conclusions also
apply to the anomalous magnetic moment. In all cases, the estimates
from the two-two method agree well with those from the two-sim method,
which we adopt as our preferred results, collected in Table~\ref{tab:Rresultsfinal}.

Recent analysis by the LHPC~\cite{Green:2012ud} shows a significant
increase in $\langle r_1^2 \rangle$ with $t_\text{sep}$, especially for their
$M_\pi=150$~MeV ensemble. We do not observe a statistically
significant effect and need data on lower-$M_\pi$ ensembles to check
the trend. The CLS-Mainz
Collaboration~\cite{Capitani:2012ef} find a dependence on $t_\text{sep}$,
however their four values of $t_\text{sep}$ are smaller than 1~fm, within the range of separations where we find excited-state contamination.

\begin{table*}
\begin{tabular}{|c|cc|cc|c|}
\hline
        & $\langle r_1^2\rangle$ ($\mbox{fm}^2$) & $\langle r_2^2\rangle$ ($\mbox{fm}^2$) & $\langle r_E^2\rangle$ ($\mbox{fm}^2$) & $\langle r_M^2\rangle$ ($\mbox{fm}^2$) & $\kappa^v$ \\\hline
310-MeV & 0.387(34)(15) & 0.474(84)(11) & 0.541(35)(10) & 0.453(67)(50) & 3.12(19)(04) \\\hline
220-MeV & 0.405(44)(17) & 0.418(84)(25) & 0.573(43)(11) & 0.415(64)(15) & 3.00(16)(01) \\\hline\hline
Extrapolation & 0.421(88)(25) & 0.368(175)(65) & 0.592(72)(21) & 0.392(109)(42) & 2.89(35)(10) \\\hline
\end{tabular}
\caption{The final results for the isovector Dirac and
Pauli charge radii, and anomalous magnetic moments.  The first errors quoted are
statistical from the overall single-elimination jackknife
procedure. The second error is an
estimate of systematic uncertainty reflecting the variation in the
estimates coming from the different fits to the form factors data.}
\label{tab:Rresultsfinal}
\end{table*}

\begin{figure}
\includegraphics[width=0.45\textwidth]{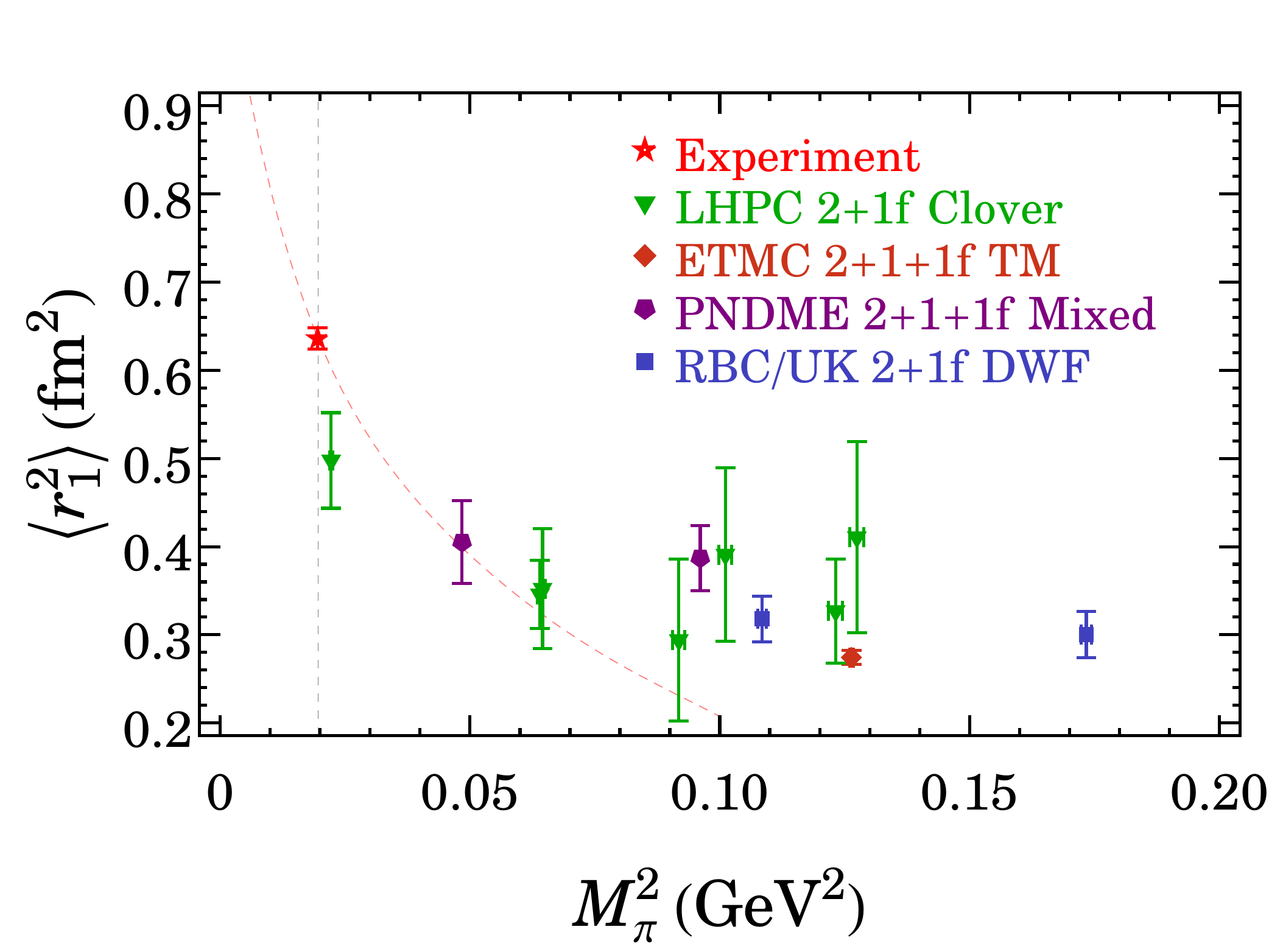}
\includegraphics[width=0.45\textwidth]{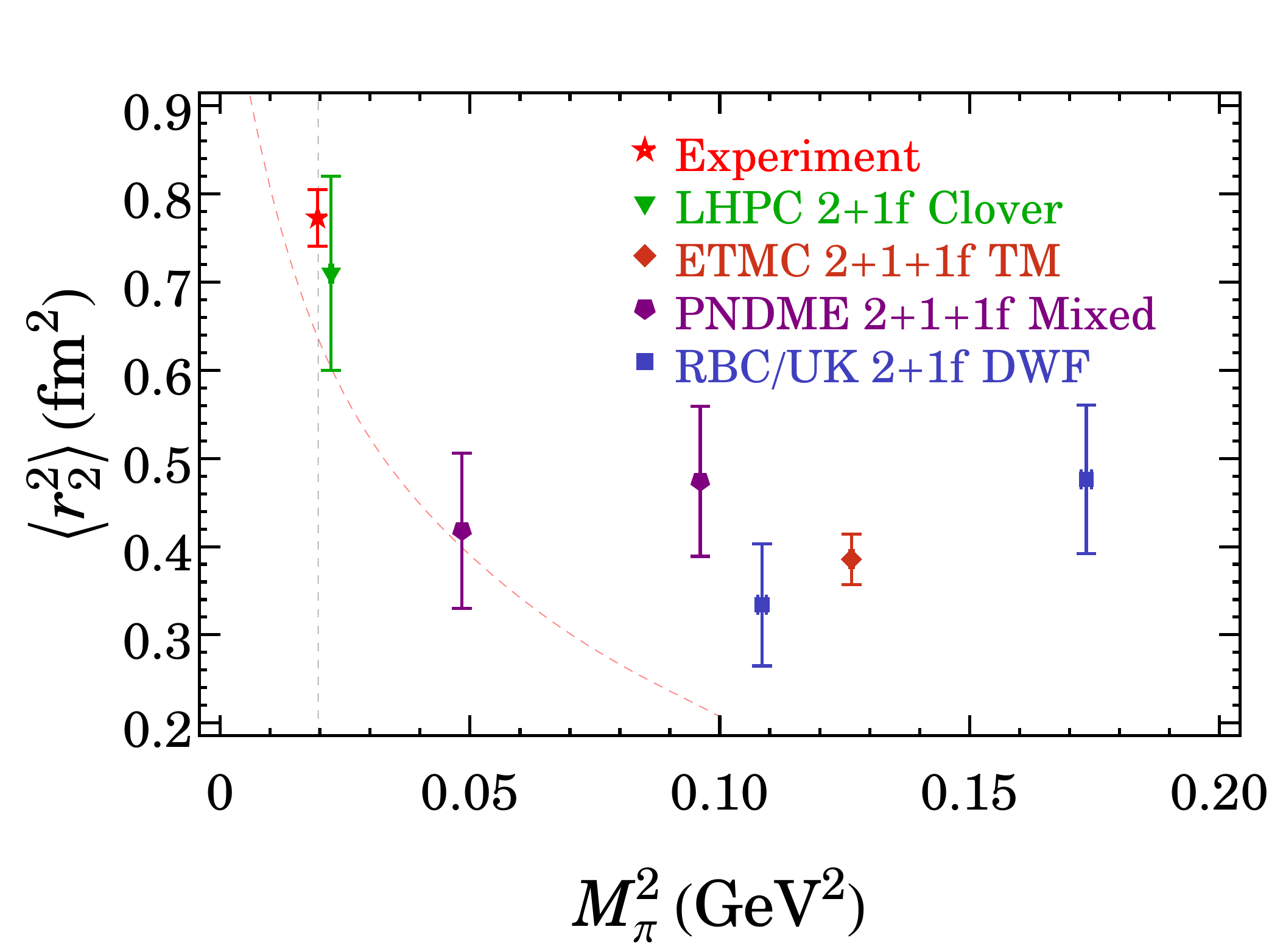}
\caption{Summary of the isovector Dirac and Pauli mean-squared radii from all
  currently existing $N_f=2+1$ and $2+1+1$ nucleon electromagnetic form-factor
  calculations~\cite{Alexandrou:2013joa,Green:2012ud,Bratt:2010jn,Yamazaki:2009zq,Syritsyn:2009mx,Lin:2008mr}.
  The dashed curve indicates the leading-order HBXPT
  prediction.}
\label{fig:all21-rv2}
\end{figure}

A summary of all the $N_f=2+1$ and $N_f=2+1+1$ lattice calculations of
the isovector Dirac and Pauli mean-squared radii
are summarized in Fig.~\ref{fig:all21-rv2} along with the
lowest-order heavy-baryon chiral perturbation theory (HBXPT) using
experimental inputs~\cite{Beg:1973sc,Bernard:1998gv}.
Note that most groups only report statistical errors, which are shown in this figure;
ours also includes the systematics due to the choice of fit-form ansatz.

\section{Conclusions}
\label{sec:end}

In this paper we demonstrate that $g_S$ and $g_T$ can be calculated to
a precision of $20\%$ or better on ensembles with $O(1000)$
configurations; we plan to increase the statistics in the
future by doubling the number of source points simulated on each lattice. This is significant, since this level of precision is
needed to leverage experimental measurements of $b$ and $b_\nu$ at the
$10^{-3}$ level to constrain novel scalar and tensor interactions at
the TeV scale.

We show that contamination from excited states can be understood and
taken into account by doing the calculations at multiple values of
$t_\text{sep}$ and performing a simultaneous fit  to all the data using
Eq.~\ref{eq:three-pt} while keeping one excited state
in the analysis.  We also find that with $O(1000)$
lattices, a consistent estimate is obtained with $t_\text{sep} \approx
1.2$~fm. In cases where sufficient computer resources are not available to
carry out studies with multiple $t_\text{sep}$, this separation should
be sufficiently large for well tuned nucleon creation/annihilation operators
at pion masses above 220~MeV.

We find that renormalization constants $Z_{A,S,T,V}$ can be calculated
with better than $5\%$ accuracy using the RI-sMOM scheme.
We do not find a window in which the data for $Z_S$
and $Z_T$ versus $q^2$ matches perturbative behavior on the
$a=0.12$~fm ensembles, however,
different estimates of individual $Z$'s and the ratios $Z_{A,S,T}/Z_V$
lie within 10\% of unity. We find that the data for the ratios
$Z_{A,S,T}/Z_V$ have less scatter than individual $Z$'s.
We assign a conservative estimate for the
systematic errors to cover the spread. Within these error estimates,
the final values of the renormalized charges obtained using different
analysis strategies give consistent results. The substantial artifacts
in the calculation of $Z$'s due to the coarse discretization of the lattice
underscore the need for a controlled continuum extrapolation of the
renormalized charges, which we plan to investigate in the future using
data at three lattice spacings ($a=0.12$, $0.09$ and $0.06$~fm).

The estimates of charges at the two values of quark masses corresponding to
$M_\pi=310$ and $220$~MeV, are in most cases within
their respective $1 \sigma$ errors. A simple linear
chiral extrapolation to the physical pion mass introduces additional
uncertainty. To reduce systematic errors due to chiral (and continuum)
extrapolation will require higher statistics simulations for at least
three values of the quark mass or simulations at the physical mass.
Since the number of configurations in the MILC ensembles are fixed,
our current strategy to increase statistics is to double the number of
source points simulated on each lattice.

\begin{acknowledgments}
We thank the MILC Collaboration for providing the 2+1+1 flavor HISQ lattices
used in our calculations. 
Simulations were carried out on computer facilities of
the USQCD Collaboration, which are funded by the Office of Science of
the U.S. Department of Energy, and by 
the Extreme Science and Engineering Discovery Environment (XSEDE), which is
supported by National Science Foundation grant number OCI-1053575.
The calculations used the Chroma software
suite~\cite{Edwards:2004sx}. TB, RG, AJ and BY are supported in part by
DOE grant No.~DE-KA-1401020. The work of HWL and SDC is supported by
DOE grant No.~DE-FG02-97ER4014. We also thank Vincenzo Cirigliano,
Alejandro Garcia and Mart\'in Gonz\'alez-Alonso  for comments and discussions,
and Jeremy Green for the updated LHPC numbers.
We acknowledge Institutional Computing at LANL for support of this project.
\end{acknowledgments}

\bibliography{pndme-a12,add}
\end{document}